\documentclass[twocolumn,revtex4,apj,iop, twocolappendix, numberedappendix]{openjournal}
\usepackage{xcolor}
\usepackage{graphicx}
\usepackage{amsfonts}
\usepackage{amssymb, bm}
\usepackage{url}
\usepackage[breaklinks,colorlinks,citecolor=blue,linkcolor=blue,urlcolor=blue]{hyperref}
\usepackage{amsmath}

\usepackage{newtxtext}
\usepackage{newtxmath}
\usepackage[normalem]{ulem}
\usepackage{fontawesome}

\renewcommand{\d}{\mathrm{d}}
\newcommand{\D}{\mathrm{D}}
\newcommand{\p}{\partial}
\newcommand{\e}{\mathrm{e}}

\renewcommand{\b}{\boldsymbol}

\renewcommand{\l}{\lambda}
\newcommand{\lo}{{\lambda_o}}

\newcommand{\dl}{\mathrm{d}_\lambda}
\newcommand{\dL}{\mathrm{d}_\Lambda}

\setcitestyle{numbers,square}
\begin{document}

\title{The Optical Expansion Scalar in Cosmological Perturbation Theory}

\author{Alexander Oestreicher$^{\;a,}$\footnote{alexo@cp3.sdu.dk}}
\author{Sofie Marie Koksbang$^{\;a}$ \vspace{0.15cm}\footnote{koksbang@cp3.sdu.dk}}

\affiliation{${}^a$ CP3-Origins, University of Southern Denmark, Campusvej 55, DK-5230 Odense M, Denmark}

\begin{abstract}
Within the geometric optics approximation, the optical expansion scalar describes the local rate of change in the amplitude of a wave or the area of a light bundle propagating through the Universe. As such, it is fundamental for describing almost all the observations we make in cosmology. It was recently pointed out that this quantity can, in principle, be reconstructed from measurements of the angular diameter distance $D_A$ and the effective observed expansion rate $\mathfrak{H}$. Motivated by this possibility, we introduce the observable $\vartheta \equiv \theta/\omega_o$, where $\theta$ is Sachs' optical expansion scalar, and $\omega_o$ the observed photon frequency. We then derive the linear perturbative expression and angular power spectrum for $\vartheta$ in a flat FLRW universe with scalar perturbations. Comparing the power spectra with those of the fluctuations in $D_A$ and with power spectra from relativistic N-body simulations reveals no immediate striking advantage of studying the fluctuations of $\vartheta$ to those of $D_A$. However, the anisotropies of $\vartheta$ do exhibit several unique qualities such as an enhanced sensitivity to the peculiar velocity and relativistic potentials. We suggest that our results may hint at $\vartheta$ being useful for developing consistency tests of cosmological observations and possibly probing departures from FLRW cosmology and modified gravity theories. The formalism developed and presented here provides the foundation for future studies into these possibilities.
\end{abstract}

\maketitle 
\tableofcontents

\section{Introduction}
Nearly everything we know about the Universe has been discovered through light beams that have travelled from their astrophysical sources to our telescopes. During the beams' propagation through spacetime, their corresponding images may be expanded, sheared and/or rotated. These changes are encapsulated in a small set of quantities, most notably the (angular diameter/luminosity) distance and the weak lensing scalars, convergence $\kappa$, and shear $\gamma$. These quantities form the basis for, or are intricate parts of, most modern cosmological probes. The clearest example is perhaps observations of type Ia supernovae which can be used to directly map the luminosity distance and thus (through Etherington's reciprocity relation \cite{Etherington_1933, Etherington_2007}) the angular diameter distance $D_A$. Other examples include Baryon Acoustic Oscillations (BAOs) \cite{Bassett_2010_BAO} which can be observed both in the galaxy distribution in the late universe and in the Cosmic Microwave Background (CMB) from the early universe. The BAOs provide a standard ruler, which can be used to directly constrain the angular diameter distance, $D_A$ \cite{Bassett_2010_BAO}. A third example is weak lensing surveys such as \cite{DES_2026_Year6_Shear_Results,KiDS_Legacy_2025_Cosmic_Shear,LSST_2012,Euclid_2011_Definition_Study_Report}, targetting galaxy ellipticities and thereby aimed at constraining the reduced shear $g=\gamma/(1-\kappa)$ \cite{Bartelmann_Schneider_2001_Weak_Lensing}. Weak lensing is e.g. a central component of Euclid \cite{Euclid_2011_Definition_Study_Report} and the Vera C. Rubin Legacy Survey of Space and Time (LSST) \cite{LSST_2012}.
\newline\indent
The quantities $D_A, D_L, \kappa$, and $\gamma$ enable us to study the cosmic expansion rate, the large scale matter distribution, and gravity at large scales \cite{Bassett_2010_BAO, Amendola_2018_Euclid_Review, Brout_2022_Pantheon+_Constraints}, demonstrating their central role in constraining our cosmological model through observations. These quantities are all determined by the evolution of a bundle of light rays following null geodesics. The evolution of such a bundle is governed by the Sachs equations \cite{Sachs_1961} and corresponding Sachs scalars: the optical expansion scalar $\theta$, and the optical shear $\sigma$. The former describes the fractional rate of change in the cross section of the light bundle, while the latter describes its anisotropic deformation, all within the geometric optics approximation (see e.g. \cite{Misner_Thorne_Wheeler_Gravitation} for an overview of the geometric optics approximation). In contrast, $\kappa$ encapsulates the total/integrated change of image size along the entire light beam, while $\gamma$ describes the total shear along the light path.
\newline\indent
As the Sachs scalars determine the evolution of any null geodesic bundle (within the geometric optics approximation), they are fundamental quantities for not only light beams, but also gravitational waves. As such, it is for example the optical expansion scalar that dictates the amplitude of both light and gravitational waves within geometric optics \cite{Misner_Thorne_Wheeler_Gravitation} and the scalar also remains important in beyond-geometric optics as seen in e.g. \cite{Darlang_2026}. The optical expansion scalar also plays a fundamental role when attempting to derive average evolution equations for the universe on the past null cone, where its average takes the role of the volume expansion rate \cite{Buchert_2023_LC_Averages}.
\newline\indent
Despite their fundamental status in geometric optics, $\theta$ and $\sigma$ have received little attention as cosmological observables in their own right. The optical scalars have instead mostly been viewed as intermediate quantities in theoretical derivations aimed at the integrated quantities $\kappa$ and $D_A$. However, as pointed out in \cite{Koksbang_2026_PRD}, $\theta$ can actually be extracted from observations by combining the angular diameter distance $D_A$, the effective observed expansion rate $\mathfrak{H}$, and redshift $z$ according to\footnote{Note that we here define $\theta$ to be half the $\theta$ used in \cite{Koksbang_2026_PRD}.}
\begin{align}\label{eq:relation}
    \theta = - \omega_o(1+z)^2\mathfrak{H}\frac{dD_A/dz}{D_A},
\end{align}
where $\omega_o$ is the photon frequency measured by the observer.
\newline\indent
This raises the natural question: {\em Can $\theta$ itself provide a useful probe of the Universe?} As data from LSST and Euclid come in, this question becomes increasingly relevant since data from these surveys will make it possible to constrain $D_A, D_A'$ and possibly $\mathcal{H}$ not only in redshift but also with angular resolution. Indeed, LSST is expected to yield $\sim 50,000$ type Ia supernovae {\em per year} of quality relevant for cosmology \cite{LSST_2009_Science_Book}, and e.g. Euclid is expected to increase the number of cosmic chronometers data greatly \cite{Moresco_2024} which can be used to constrain $\mathcal{H}$ in both angular position and redshift. In addition, several missions such as Euclid  are greatly increasing our amount of BAO data \cite{Euclid_2011_Definition_Study_Report} which can be used to constrain $\mathcal{H}$ as a function of redshift, although smoothed over the sky/angular position for common treatments of BAO data.
\newline\indent
Although $\theta$ will be harder to constrain than the individual components needed to measure it, it may offer a more direct window into the fundamentals of light propagation. Naturally, an analysis of $\theta$ cannot provide more information than a full joint analysis of the observables needed to reconstruct it. Nonetheless, its status as a fundamental variable in the theory of light propagation and its close relation to the Ricci curvature may mean that this particular combination of data is particularly sensitive to the properties of the underlying space time. The optical expansion scalar might therefore e.g. provide new insights into the theory of gravity and/or cosmological models. $\theta$ has the advantage of constraining the local rate of expansion at every point along a light ray rather then the integrated signal that $D_A$ and $\kappa$ constrain. As light path-integrated observables, $\kappa$ and $D_A$ inevitably mix information from different scales and cosmic epochs, which limits their ability to retain detailed scale- and time-dependent information. For instance, projection effects in the convergence field suppresses BAO features which can only be reconstructed through complicated tomography-based techniques \cite{Touzeau_2026}. 
\newline\indent
However, it is currently unclear whether $\theta$ offers genuinely new insights compared to $\kappa$ and $D_A$. In particular, we note that although $\theta$ is inherently locally {\em defined}, it is not locally {\em determined} since we make our observations on the past light cone and must solve the Sachs equations \cite{Sachs_1961} to obtain its value at a given point along a light path. As we will see later, this means that the expression for $\theta$ includes terms integrated along the light path, similar to those appearing in the expressions for $\kappa$ and $D_A$. 
\newline\indent
To learn whether $\theta$ offers genuinely new insight, we must first establish the theoretical framework for studying $\theta$: We need to understand how it behaves in a realistic, cosmological spacetime, how it relates to metric perturbations and what statistical signatures it is expected to exhibit. Our primary goal in this paper is thus to establish the theoretical foundation required for studying $\theta$ as a cosmological observable. We derive its perturbative expression, identifying its monopole and dipole, and compute its angular power spectrum. To verify the validity of our perturbative results and characterize the signal in realistic settings, we further compare our analytical results with results from a relativistic cosmological simulation obtained with the simulation code \texttt{gevolution} \cite{Adamek2016_Nat}. We furthermore compare the fluctuations in $\theta$ to those in $D_A$ which at lowest order in perturbation theory are equal to $\kappa$.
\newline\newline
The paper is organized as follows: In Sect.~\ref{sect:theoretical_background}, we recap the fundamentals of light propagation, explaining how $\theta$ may be constrained observationally and suggesting the observable $\vartheta=\theta/\omega_o$. In Sect.~\ref{sect:FLRW} we show results for $\vartheta$ in a perfect FLRW universe. We then derive the expression for $\vartheta$ to first order in perturbation theory in Sect.~\ref{sect:perturbations}. Using these results we derive the angular power spectra for $\vartheta$ in Sect.~\ref{sect:power_spectra}, before showing numerical results for these power spectra and comparing them to simulation results in Sect.~\ref{sect:numerical_results}. We conclude in Sect.~\ref{sect:conclusions}.

\section{Theoretical Background}
\label{sect:theoretical_background}
In this section, we recap the basic formalism used to describe light beams and show how the relation \eqref{eq:relation} is obtained in detail. We finish the section by defining a new observable, $\vartheta$.

\subsection{The Sachs Equations}
We will start by briefly recapping the theory of light bundles and introduce the Sachs optical scalars. We are interested in an observer receiving a bundle of light rays from a distant astronomical source. We assume this bundle to be infinitesimal and converge at the observer's location. The observer makes observations in their local Lorentz frame which we can describe using an observer tetrad $\{\hat e^\mu_a\}$, with $\hat e^\mu_0=u^\mu$ the observer's 4-velocity \cite{Mitsou_Yoo_2020_Tetrad_Formalism_for_Cosmology}. We denote vectors in the tetrad with Latin indices from the beginning of the alphabet and spacetime indices with Greek indices. When there is chance of confusion, quantities belonging to the observer in their own frame are additionally denoted with a hat. The observer measures the frequency 
\begin{align}
    \label{eq:def_omega}
    \omega \equiv \hat k^0 = \hat e^0_\mu k^\mu=-u_\mu k^\mu\;, 
\end{align}
and sees the light coming from the direction 
\begin{equation}
    \label{eq:def_hat_ei}
    e^i \equiv -\frac{\hat k^i}{\hat k^0}=-\frac{\hat e^i_\mu k^\mu}{\omega}\;.
\end{equation}
Here $k^\mu$ is the spacetime wave vector and the direction is divided by the frequency to ensure that it is a unit vector and includes a minus sign so that it points from the observer to the source. In order to describe bundles of light rays, we now introduce two basis vectors $\hat s^i_A$, $A\in{1,2}$,  perpendicular to $e^i$, spanning a screen in the observer's sky. These two vectors form the so called Sachs Basis. We can introduce spacetime versions of $e^i$ and $\hat s_ A^i$ according to 
\begin{equation}
    d^\mu = e^i\hat e_i^\mu\;, \qquad s_A^\mu = \hat s^i_A\hat e^\mu_i\;.
\end{equation}
In order to sensibly compare quantities in the Sachs basis at different points along a fiducial light ray in the light beam, we require that the basis vectors are parallel transported along the ray, i.e. $k^\nu\nabla_\nu s_A^\mu=0$. With these definitions it is possible to decompose the wave vector $k^\mu$ into a part parallel to the 4-velocity and the spatial direction vector $d^\mu$ 
\begin{equation}
    \label{eq:k_decomp}
    k^\mu = \omega (u^\mu-d^\mu)\;.
\end{equation}
By definition, $d^\mu u_\mu=0$. 
\\ \\
We can now describe a bundle of light rays as a set of curves $\gamma^\mu(\Lambda,\b \alpha)$ parametrized by an affine parameter $\Lambda$ and two further parameters $\b\alpha=(\alpha_1,\alpha_2)$ labelling the individual rays, which can be thought of as angular coordinates on the observer's sky. We study the separation vector between two neighbouring rays at constant $\Lambda$
\begin{equation}
    \label{eq:xi_definition}
    \xi^\mu \equiv \gamma^\mu(\Lambda, \b\alpha+\delta\b\alpha)-\gamma^\mu(\Lambda,\b\alpha)\;,
\end{equation}
projected onto the Sachs basis
\begin{equation}
    \xi^A\equiv\xi^\mu s_\mu^A\;.
\end{equation}
The change in $\xi^A$ along the null geodesic tells us how the light bundle evolves, and is governed by the equation \cite{Seitz_1994}
\begin{equation}
    \label{eq:xiA_sachs_evolution}
    \D_\Lambda\xi^A=S^A{}_B\xi^B\;.
\end{equation}
Here, $\D_\Lambda$ denotes a directional derivative along the affine parameter $\Lambda$ defined as $\D_\Lambda \equiv k^\mu\nabla_\mu$ and we introduced the deformation rate matrix 
\begin{equation}
    S_{AB}=s_A^\nu s_B^\mu \nabla_\mu k_\nu\;.
\end{equation}
It is a symmetric matrix and can thus be decomposed into its trace and a trace free symmetric matrix as 
\begin{equation}\label{eq:decomp_S}
    \b S = \begin{pmatrix}
        \theta & 0 \\ 0 & \theta
    \end{pmatrix} 
    +
    \begin{pmatrix}
        -\sigma_1 & \sigma_2 \\
        \sigma_2 & \sigma_1
    \end{pmatrix}\;.
\end{equation}
The scalars $\theta$, $\sigma_1$ and $\sigma_2$ are known as the optical expansion and shear scalars due to there respective action on the shape of the light beam. The optical expansion scalar $\theta$ causes an isotropic expansion or contraction of the beam while $\sigma_1$ and $\sigma_2$ shear the shape of the beam causing a initial round image to appear elliptical. It is common to combine the shear into the single complex quantity $\sigma=\sigma_1+i\sigma_2$. The evolution of $\theta$ and $\sigma$ is governed by the two Sachs equations \cite{Sachs_1961}
\begin{align}
    \label{eq:sachs1}
    \d_\Lambda\theta+\theta^2+\vert\sigma\vert^2&=\mathcal{R}\;, \\ 
    \label{eq:sachs2}
    \d_\Lambda\sigma+2\theta\sigma&=\mathcal{W}\;.
\end{align}
Here, the directional derivative has reduced to $\d_\Lambda=k^\mu\p_\mu$ because it is acting on scalars. $\mathcal{R}$ and $\mathcal{W}$ are the Ricci and Weyl lensing scalars defined as
\begin{align}
    \mathcal{R}&=-\frac{1}{2}R_{\mu\nu} k^\mu k^\nu\;, \\
    \mathcal{W}&=C_{\mu\kappa\sigma\nu}(s^\mu_1-i s^\mu_2)k^\kappa k^\sigma (s^\nu_1-i s^\nu_2)\;,
\end{align}
where $R_{\mu\nu}$ is the Ricci tensor and $C_{\mu\kappa\sigma\nu}$ the Weyl tensor. The optical expansion scalar is related to the area of the light bundle $\mathrm{A}$ and thereby the angular diameter distance $D_A$ as \cite{Schneider_Ehlers_Falso_1992}
\begin{equation}\label{eq:theta_DA}
	\theta = \d_\Lambda \ln\mathrm{A}=\d_\Lambda\ln D_A\;.
\end{equation}
Before studying the optical expansion scalar $\theta$ in more detail and discussing how we can constrain it observationally in Sect.~\ref{sect:vartheta_def}, we first discuss how the optical scalars are related to the angular diameter distance and the usual weak lensing observables in the next subsection.

\subsection{The Jacobi Matrix}
Taking as second derivative of \eqref{eq:xiA_sachs_evolution} one finds the evolution equation \cite{Seitz_1994}
\begin{equation}
    \label{eq:sachs_vector_evolution}
    \D_\Lambda^2\xi^A=\mathcal{R}^A{}_B\,\xi^B\;.
\end{equation}
The matrix $\b{\mathcal{R}}$ is known as the optical tidal matrix and can be written as 
\begin{equation}
    \b{\mathcal{R}}=\begin{pmatrix}
        \mathcal{R} & 0 \\
        0 & \mathcal{R}
    \end{pmatrix}
    + \begin{pmatrix}
        -\mathrm{Re}(\mathcal{W}) & \mathrm{Im}(\mathcal{W}) \\
        \mathrm{Im}(\mathcal{W}) & \mathrm{Re}(\mathcal{W})
    \end{pmatrix}\;.
\end{equation}
Since \eqref{eq:sachs_vector_evolution} is a linear equation there exists a linear map between the initial conditions at the observer and any other point along the ray. As we chose a beam that converges at the observer $\xi^A_o=0$ and we have 
\begin{equation}
    \xi^A=\mathcal{D}^A{}_B\theta^A_o\;,
\end{equation}
where $\theta^A_o=\omega_o^{-1} \dL\xi^A\vert_o$ is the observed angular separation in the observer's sky. The matrix $\mathcal{D}$ is called the Jacobi map and fulfils the propagation equation
\begin{equation}
    \label{eq:jacobi_transport}
    \mathrm{D}_\Lambda^2 \b{\mathcal{D}}=\b{\mathcal{R}\mathcal{D}}.
\end{equation}
The angular diameter distance is defined through the ratio of the physical size $A_s$ and the observed angular size $\Omega_o$ of an astronomical source and is as such given by the determinant of the Jacobi map
\begin{equation}
    D_A \equiv \sqrt{\frac{A_S}{\Omega_o}}=\sqrt{\det(\b{\mathcal{D}})}\;.
\end{equation}
One may introduce the so-called amplification matrix 
\begin{equation}
    \b{\mathcal{D}}=\b{\mathcal{A}\bar{\mathcal{D}}}\;,
\end{equation}
where $\b{\bar{\mathcal{D}}}=\bar D_A\b 1_2$ is the monopole/sky-average Jacobi map. This amplification matrix can be decomposed into its trace, a traceless symmetric matrix, and a traceless anti-symmetric matrix, with components $\kappa$, $\gamma_1$, $\gamma_2$,  and $\psi$, according to 
\begin{equation}
    \b{\mathcal{A}}= \begin{pmatrix}
        1-\kappa-\gamma_1 & \gamma_2-\psi \\
        \gamma_2+\psi & 1-\kappa+\gamma_1
    \end{pmatrix}\;.
\end{equation}
The scalar $\kappa$ is the weak lensing convergence, $\gamma_1$ and $\gamma_2$ the shear scalars, and $\psi$ the rotation due to their effect on the image of an astronomical source \cite{Fleury_2015_PhD}. Within linear perturbation theory, the weak lensing convergence $\kappa$ and the angular diameter distance are related as 
\begin{equation}
    \label{eq:kappa=-deltaDA}
    \kappa = \frac{\bar D_A -D_A}{\bar D_a}\simeq -\frac{\delta D_A}{\bar D_A}\;,
\end{equation}
where $\bar D_A$ denotes the monopole/sky-average while $\delta D_A$ is the first order perturbation to the angular diameter distance. 
The deformation rate matrix and the Jacobi map are related according to \cite{Seitz_1994}
\begin{equation}
    \label{eq:rel_D_S}
    \D_\Lambda\b{\mathcal{D}}= \b{S\mathcal{D}}\;.
\end{equation}
The Jacobi map $\b{\mathcal{D}}$ (or equivalently $D_A, \kappa,\gamma$ and $\psi$) tells us how much the separation vector/image of the astronomical source has changed along the entire light path. The deformation rate matrix $\b{\mathcal{S}}$ (or equivalently, $\theta$ and $\sigma$) tells us the local rate of change of the same light bundle at every point along the light path. 

\subsection{Definition of $\vartheta$}
\label{sect:vartheta_def}
Since the affine parameter $\Lambda$, is not an observable, we rewrite \eqref{eq:theta_DA} in terms of redshift derivatives. To achieve this, we first remember that the redshift along a light ray parametrized by $\Lambda$ is given by
\begin{equation}
    \label{eq:def_redshift}
    1+z(\Lambda)=\frac{\omega(\Lambda)}{\omega_o}\;,
\end{equation}
where $\omega_o$ denotes the frequency of the light at the observer. Taking a derivative of this expression with respect to the affine parameter and using \eqref{eq:def_omega} leads to
\begin{align}
    \D_\Lambda z&=\frac{1}{\omega_o}\D_\Lambda\omega=-\frac{1}{\omega_o}k^\mu\nabla_\mu (k^\nu u_\nu)=-\frac{1}{\omega_o}k^\mu k^\nu\nabla_\mu u_v\;.
\end{align}
The gradient of the 4-velocity appearing here can be decomposed as follows \cite{Relativistic_Cosmology_Book}
\begin{equation}
    \nabla_\mu u_\nu = \frac{1}{3}\theta^{(u)} (g_{\mu\nu}+u_\mu u_\nu)+\sigma^{(u)}_{\mu\nu}+\omega^{(u)}_{\mu\nu}-u_\mu A^{(u)}_\nu\;,
\end{equation}
where $\theta^{(u)}$, $\sigma^{(u)}$, $\omega^{(u)}$ are the fluid expansion, shear, and rotation rate, and $A_\mu^{(u)}$ is the fluid acceleration. The shear, rotation, and acceleration are all orthogonal to the fluid flow $u^\mu$ and the rotation is additionally an anti-symmetric matrix. With this, and the decomposition of the wave vector \eqref{eq:k_decomp} it quickly follows that 
\begin{equation}
\label{eq:gen_hubble_rate}
    \D_\Lambda z= -\frac{\omega^2}{\omega_o} \left(\frac{1}{3}\theta^{(u)}+d^\mu d^\nu\sigma^{(u)}_{\mu\nu}-d^\mu A^{(u)}_\mu\right)= -\frac{\omega^2}{\omega_o}\mathfrak{H}\,
\end{equation}
where we identified the effective observed Hubble rate $\mathfrak
{H}$. $\mathfrak{H}$ appears as the lowest order term in a series expansion of observed distances around the observer and as such generalizes the Hubble law (at low redshifts) in arbitrary spacetimes \cite{Clarkson_2000PhD,Umeh_2013_PhD,Heinesen_2021_DL_decomposition}. If the fluid flow is geodesic, the acceleration term vanishes and $\mathfrak{H}$ reduces to the generalized Hubble rate governing the rate of change of neighbouring galaxies in general spacetimes \cite{Relativistic_Cosmology_Book}. With this, we can now replace the derivative with respect to the affine parameter in \eqref{eq:theta_DA} to find
\begin{equation}
	\theta = -\omega_o(1+z)^2\mathfrak{H}\frac{1}{D_A}\frac{\d D_A}{\d z}\;.
\end{equation}
All quantities on the right hand side can be constrained from observations without relying on a specific cosmological model as demonstrated in \cite{Koksbang_2026_PRD}.
\newline\indent
Within the geometric optics approximation, observational results are frequency-independent. Thus, although we naturally observe in a specific and well-known frequency interval for any given observation, we may consider the frequency a matter of a rescaling of results. We therefore choose to divide it out and define the new observable 
\begin{equation}\label{eq:vartheta}
	\vartheta \equiv \frac{\theta}{\omega_o}=-(1+z)^2\mathfrak{H}\frac{1}{D_A}\frac{\d D_A}{\d z}\;.
\end{equation}
Note that all we have assumed in order to obtain the above relation is that spacetime is a 4D Lorentzian manifold, that the geometric optics approximation holds (so that light follows null geodesics), and that the mapping between the affine parameter and redshift is one-to-one.

\section{FLRW Limit}
\label{sect:FLRW}
\begin{figure}
    \centering
    \includegraphics[width=0.8\linewidth]{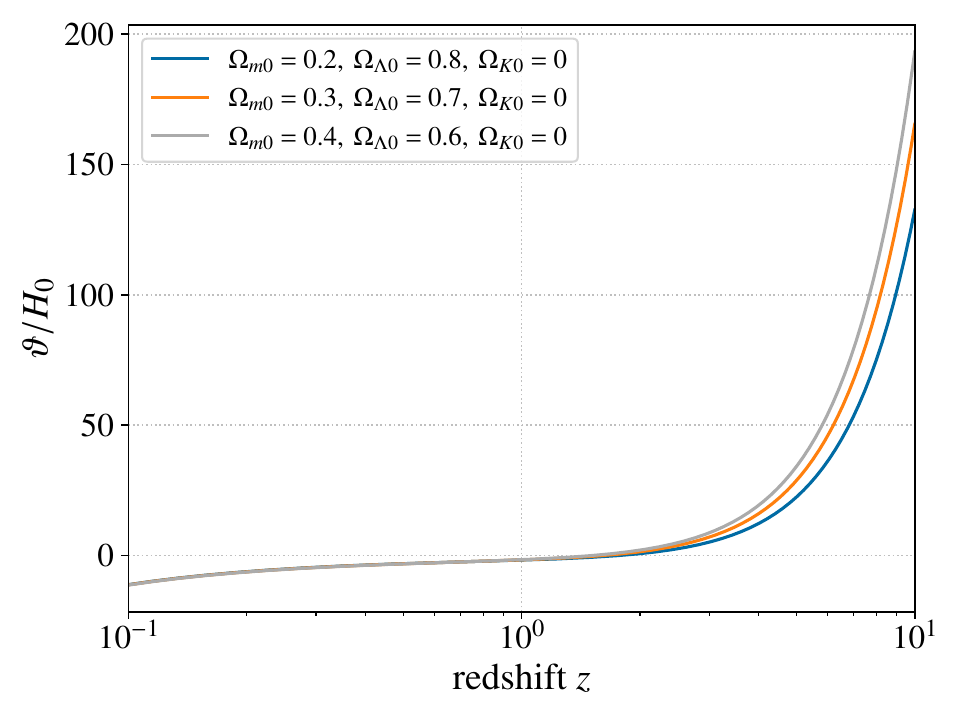}
    \includegraphics[width=0.8\linewidth]{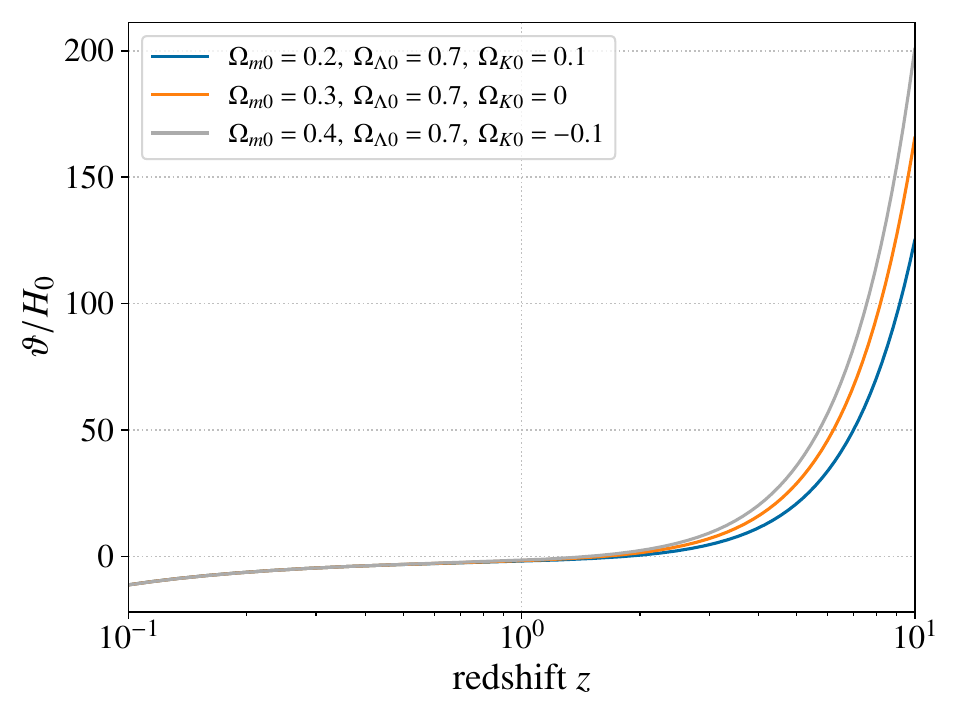}
    \caption{$\vartheta/H_0$ for three $\Lambda$CDM+curvature models. The model parameters are indicated in the legends.}
    \label{fig:monopole}
\end{figure}

Before moving on to derive the perturbative expression for $\vartheta$, we will consider what it reduces to in a Friedmann-Lemaitre-Robertson-Walker (FLRW) spacetime with line element
\begin{equation}
    \d s^2 = a^2 \left[-\d\eta^2 + \d r^2 + f^2_K(r)\d\Omega^2\right]\;,
\end{equation}
where 
\begin{equation}
    f_K(r)=\begin{cases}
        K^{-1/2}\sin(K^{1/2}r) & (K> 0) \\ 
        r & (K=0)\quad\;. \\
        \vert K\vert^{-1/2}\sinh(\vert K\vert^{1/2}r) & (K<0)
    \end{cases}
\end{equation}
Here $\eta$ is conformal time, $r$ is the co-moving radial distance, $\d\Omega$ the usual infinitesimal solid angle element, $K$ the spatial curvature, $a$ the scale factor, and we have set $c=1$. We use the usual convention of setting $a_o=1$. For this metric the angular diameter distance is simply
\begin{equation}
    D_A=af_K(r)
\end{equation}
and a straightforward calculation shows that it's derivative is
\begin{align}
    \frac{\d D_A}{\d z} &= \frac{\d a}{\d z}f_K(r)+a\frac{\d r}{\d z}\frac{\d f_K}{\d r} \nonumber \\ 
    &=-a^2f_k(r)+a^2(\mathcal{H})^{-1}\sqrt{1-Kf_K^2(r)}\;,
\end{align}
where we introduced the conformal Hubble parameter $\mathcal{H}=\dot a/a$. Dots denote derivatives with respect to conformal time $\eta$. In FLRW models, the generalized Hubble rate $\mathfrak{H}$ reduces to the usual expression for the Hubble parameter $\mathfrak{H}=H=a^{-1}\mathcal{H}$. Plugging the above into \eqref{eq:vartheta} we find that the FLRW limit of $\vartheta$ can be written as
\begin{align}
	\label{eq:theta_rw}
    \vartheta(z)&=(1+z)^2\mathcal{H}-(1+z)^2\sqrt{f_k^{-2}(r)-K} \\ \nonumber 
    &=(1+z)^2\mathcal{H}-(1+z)\sqrt{D_A^{-2}-K(1+z)^2}\;.
\end{align}
We plot the FLRW limit of $\vartheta$ in Fig.~\ref{fig:monopole} for a few different $\Lambda$CDM+curvature models. To make the results dimensionless, we show $\vartheta/H_0$, where $H_0$ is the present day value of the Hubble function of the given model. As seen, the FLRW limit of $\vartheta$ exhibits only a modest dependence on FLRW cosmological parameters, in particular at very low redshifts. Moreover, these parameters are already constrained by the independent measurements of $H$ and $D_A$ needed to infer $\vartheta$. Consequently, the primary usefulness of $\vartheta$ most likely lie elsewhere. Specifically, combining $H$ and $D_A$ into $\vartheta$ provides a more direct probe of the underlying spacetime, as it grants access to one of the dynamical quantities governing the evolution of light bundles. In this context, \eqref{eq:theta_rw} is perhaps best viewed as an FLRW consistency test. Indeed, solving \eqref{eq:theta_rw} for $K$ one finds
\begin{equation}
    K = \frac{D_A^{-2}-H^2\left(1+\frac{\vartheta^2}{(1+z)^2H^2}-\frac{2\vartheta}{1+z}\right)}{(1+z)^2}\;,
\end{equation}
which is the FLRW limit of the integrated version of the Clarkson-Bassett-Lu test \cite{Clarkson_2008_CBL} (see eq.~(6) in \cite{Koksbang_2026_PRL}\footnote{Note that \cite{Koksbang_2026_PRL} define $\hat \theta = \theta/2$ and set $\omega_o=1$.}).

\section{Perturbations}
\label{sect:perturbations}
We now proceed to derive the expression for $\vartheta$ in a flat FLRW universe with scalar perturbations.
 
\subsection{Assumptions and Setup}
We will work in the Newtonian gauge, where the line element can be written as
\begin{align}\label{eq:metric}
	\widetilde{\d s}^2 &= \tilde g_{\mu\nu}\d x^\mu \d x^\nu \nonumber \\ 
    &= a^2(\eta) \left[-(1+2\Psi)\d\eta^2+(1-2\Phi)\delta_{ij}\d x^i\d x^j\right]\;.
\end{align}
$\Psi$ and $\Phi$ are the two Bardeen potentials \cite{Bardeen_1980} and $x^i$ are co-moving coordinates. Time derivatives with respect to $\eta$ will be denoted with a dot as earlier.
\newline\indent
Light paths are invariant under conformal transformations \cite{Fleury_2015_PhD} and we will therefore for simplicity perform our calculations using the conformally rescaled metric 
\begin{equation}
	g_{\mu\nu} = a^{-2}\tilde g_{\mu\nu}.
\end{equation}
We split the background wave vector into an amplitude and unit direction vector according to
\begin{equation}
	\bar k^\mu = \bar k^0(1, n^i)\;.
\end{equation}
From the null geodesic equation for the background it follows that $\bar k^0$, $\bar k^i$ and $n^i$ are constant along the light path, as determined by \eqref{eq:null_geodesic_eq_bar_k}. We can therefore use the affine freedom to set $\bar k^0=1$ everywhere along the light path. We will now consider a central light ray in both the physical and background light beams. The background light path is denoted by $\bar x^\mu$. The background light path deviates from the full ray $x^\mu$ by a small perturbation $\delta x^\mu\equiv x^\mu-\bar x^\mu$. The background and physical light paths have tangent vectors $\bar k^\mu$ and $k^\mu$, connected by $\delta k^\mu\equiv k^\mu-\bar k^\mu$. We parametrize the rays with an affine parameter $\lambda$ along the background trajectory such that 
\begin{equation}
	\bar k^\mu = \frac{\d\bar x^\mu}{\d\lambda}
\end{equation}
and 
\begin{equation}
	x^\mu(\lambda)=\bar x^\mu(\lambda)+\delta x^\mu(\lambda)\;.
\end{equation}
The directional derivative along $\lambda$ is 
\begin{equation}
	\dl = \bar k^\mu \p_\mu\;.
\end{equation}
We also introduce the affine parameter $\Lambda$ along $x^\mu$, such that the directional derivative along the full path is 
\begin{equation}
    \label{eq:deriv_pertb}
	\dL = k^\mu\p_\mu = \dl + \delta k^\mu\p_\mu\;.
\end{equation}
Having set $\bar k^0=1$, the affine parameter $\lambda$ along the background trajectory is related to the background conformal time $\eta$ and co-moving distances $r$ according to
\begin{equation}
	\label{eq:dl=deta=-dr_theta}
	\d\lambda = \d\eta =-\d r\;.
\end{equation}
To declutter notation, we will denote integrals along the past light cone from the observer to the source as
\begin{equation}
	\int_\lambda f(\lambda') \equiv \int_{\lambda_o}^{\lambda}\d\lambda'f(\lambda')\;.
\end{equation}
Throughout the derivation we frequently need the background optical expansion scalar and angular diameter distance. Reading them off from the results in Sect.~\ref{sect:FLRW}, with $a=1$ and $K=0$, these can simply be written as
\begin{align}
    \label{eq:background_DA}
    \bar D_A & = r = \lambda_o-\lambda=\eta_o-\eta \;, \\
    \label{eq:background_theta}
    \bar\theta &=-\frac{1}{\bar D_A}\;.
\end{align}
With this in place, we proceed to derive the perturbative expression for $\vartheta$.

\subsection{Perturbative Expression for $\vartheta$}
The onset of our derivation is the Sachs equation \eqref{eq:sachs1} determining the evolution of $\theta$ along a light path. In an FLRW spacetime there is no shear, as this would indicate a preferred direction, and since $\sigma$ appears only squared in the first Sachs equation it can therefore not contribute to $\theta$ at first order in perturbation theory. We therefore have
\begin{equation}
	\label{eq:sachs1_no_shear}
	\dL\theta+\theta^2=\mathcal{R}\;.
\end{equation}
We expand both $\theta$ and $\mathcal{R}$ to first order and write
\begin{equation}
	\theta = \bar\theta+\delta\theta\;,\qquad \mathcal{R} = \bar{\mathcal{R}}+\delta\mathcal{R}\;.
\end{equation}
At first order, and subtracting the background, \eqref{eq:sachs1_no_shear} becomes
\begin{equation}\label{eq:diff_eq_dtheta}
	\dl\delta\theta+\delta k^\mu\p_\mu\bar\theta+2\bar\theta\delta\theta = \delta\mathcal{R}\:,
\end{equation}
where we used \eqref{eq:deriv_pertb} to replace the directional derivative along the physical path $\d_\Lambda$ with that along the background path $\d_\lambda$. Using the relation \eqref{eq:theta_DA}, we see that this can be rewritten as
\begin{align}
	\label{eq:different_eq_dtheta}
	\dl(\bar D_A^2\delta\theta)&=\bar D_A^2(\delta\mathcal{R}-\delta k^\mu\p_\mu\bar\theta)\;, 
\end{align}
which can be integrated to yield
\begin{align}
	\label{eq:dtheta_start}
	\delta\theta(\lambda) &= \frac{1}{\bar D_A^2} \int_\l\bar D_A^2(\delta\mathcal{R}-\delta k^\mu\p_\mu\bar\theta) \nonumber \\ 
    &= \frac{1}{\bar D_A^2} \int_\l(\bar D_A^2\delta\mathcal{R}-\delta k^0)\;.
\end{align}
We here used that the background angular diameter distance, $\bar D_A$, vanishes at the observer and evaluated the last term in the first line using the expressions \eqref{eq:background_DA} and \eqref{eq:background_theta}. To continue, we now need the perturbations to the Ricci lensing scalar $\delta\mathcal{R}$ as well as those of the time component of the wave vector $\delta k^0$. We derive them in Appx.~\ref{appx:geometry}, finding 
\begin{align}
	\label{eq:dR_theta}
	\delta\mathcal{R}= -\frac{1}{2}\left[\nabla_\perp^2(\Phi+\Psi)+2\dl^2\Phi-\frac{2}{r}\dl(\Phi+\Psi)+\frac{2}{r}(\dot\Phi+\dot\Psi)\right]
\end{align}
and 
\begin{equation}
	\label{eq:dk0_theta}
	\delta k^0(\lambda)=\delta k^0(\lambda_o)-2[\Psi]_\lo^\l+\int_\l\,(\dot\Phi+\dot\Psi)\;.
\end{equation}
We here expressed $\delta\mathcal{R}$ using the perpendicular gradient $\nabla_\perp=\nabla^2-n^i n^j\p_i\p_j+\frac{2}{r}n^i\p_i$, commonly used to express the lensing potential \cite{Bonvin_2008_pec_motion_weak_lensing,Umeh_2014_DA_second_order_derivation}. Before inserting these expressions into \eqref{eq:dtheta_start}, let us consider the two terms in \eqref{eq:dR_theta} that contain a total derivative. Inserting only these into \eqref{eq:dtheta_start} and using that $\bar D_A=\lambda_o-\lambda$ and therefore $\dl \bar D_A=-1$, partial integration yields 
\begin{align}
	&\,-\frac{1}{2}\frac{1}{\bar D_A^2} \int_\l\bar D_A^2\left[2\dl^2\Phi-\frac{2}{r}\dl(\Phi+\Psi)\right] \nonumber \\
	&=-\dl\Phi-\frac{1}{\bar D_A^2} \int_\l\bar D_A\left[2\dl\Phi-\dl(\Phi+\Psi)\right]\nonumber \\
	&=-\dl\Phi-\frac{1}{\bar D_A}(\Phi-\Psi)-\frac{1}{\bar D_A^2}\int_\l(\Phi-\Psi)\;.
\end{align}
All terms together, we have
\begin{align}
	\delta\theta(\lambda) =&\,-\dl\Phi-\frac{1}{\bar D_A}(\Phi-\Psi)-\frac{1}{\bar D_A^2}\int_\l(\Phi-\Psi) \nonumber \\ 
	&\,-\frac{1}{2\bar D_A^2}\int_\l \bar D_A^2\left[\nabla_\perp^2(\Phi+\Psi)+\frac{2}{r}(\dot\Phi+\dot\Psi)\right] \nonumber \\
	&\,+\frac{1}{\bar D_A^2}\int_\l\left(\delta k^0(\lambda_o)-2[\Psi]_\lo^{\l'}+\int_{\l'}\,(\dot\Phi+\dot\Psi)\right)\;.
\end{align}
Integrating over the two observer terms $\delta k^0_o$ and $\Psi_o$ and again using $\bar D_A = \lo-\l$ this is equivalent to 
\begin{align}
	\delta\theta(\lambda) =&\,-\dl\Phi-\frac{1}{\bar D_A}(\Phi-\Psi)-\frac{1}{\bar D_A^2}\int_\l(\Phi-\Psi) \nonumber \\ 
	&\,-\frac{1}{2\bar D_A^2}\int_\l \bar D_A^2\nabla_\perp^2(\Phi+\Psi)-\frac{1}{\bar D_A^2}\int_\l \bar D_A(\dot\Phi+\dot\Psi) \nonumber \\
	&\,-\frac{1}{\bar D_A}(\delta k^0_o+2\Psi_o)-\frac{2}{\bar D_A^2}\int_\l\Psi \nonumber \\
    &\,+\frac{1}{\bar D_A^2}\int_\l\int_{\l'}\,(\dot\Phi+\dot\Psi)\;.
\end{align}
Inserting $\bar D_A=r$, $\d\lambda=-\d r$ and $\dl=\p_0+n^i\p_i$, and interpreting the entire expression as a function of the background conformal time, leaves us with
\begin{align}
	\delta\theta(\eta) =&\,-\dot\Phi-n^i\p_i\Phi-\frac{1}{r}(\Phi-\Psi)+\frac{1}{r^2}\int_r(\Phi-\Psi) \nonumber \\ 
	&\,+\frac{1}{2r^2}\int_r r'^2\nabla_\perp^2(\Phi+\Psi)+\frac{1}{r^2}\int_r r'(\dot\Phi+\dot\Psi) \nonumber \\
	&\,-\frac{1}{r}(\delta k^0_o+2\Psi_o)+\frac{2}{r^2}\int_r\Psi+\frac{1}{r^2}\int_r\int_{r'}\,(\dot\Phi+\dot\Psi)\;.
\end{align}
We will now convert the result to the expanding FLRW spacetime described by the metric $\tilde g_{\mu\nu}$, noting that all the quantities appearing on the right hand side of the equation above are the same in the conformally rescaled spacetime and the expanding spacetime. We also divide by the frequency at the observer $\omega_o$. The angular diameter distance, affine parameter and frequency transform as $\tilde D_A = a D_A$, $\d\tilde\Lambda=a^2\d\Lambda$, and $\tilde\omega=\omega/a$ when going between $\tilde g_{\mu\nu}$ and $g_{\mu\nu}$ (see e.g. \cite{Fleury_2015_PhD} Table 5.1). From this, we find
\begin{equation}
	 \tilde \vartheta = \frac{\tilde\theta}{\tilde\omega_o}= \frac{1}{\tilde\omega_o}\frac{1}{\tilde D_A}\frac{\d\tilde D_A}{\d\tilde\Lambda}=\frac{1}{\omega_o}\left(a^{-3}\dL a +a^{-2}\theta\right)\;,
\end{equation}
where we used that $a_o=1$. Expanding to first order in perturbation theory, using \eqref{eq:deriv_pertb} and our choice $\bar k^0=\bar\omega=1$, we find 
\begin{align}
	\delta\tilde\vartheta&=a^{-2}\left(a^{-1}\delta k^\mu\p_\mu a+\delta\theta-(a^{-1}\dl a+\bar\theta)\delta\omega_o\right) \nonumber \\
	&=a^{-2} \left(\mathcal{H}\delta k^0+\delta\theta-\left(\mathcal{H}-\frac{1}{r}\right)\delta\omega_o\right)\;.
\end{align}
Inserting our result for $\delta\theta$ and the expression \eqref{eq:dk0_theta} for $\delta k^0$ we find
\begin{align}
	\delta\tilde\vartheta(\eta) =&\, a^{-2}\Bigg[\left(\mathcal{H}-\frac{1}{r}\right)(\delta k^0_o-\delta \omega_o+2\Psi_o) \nonumber \\
    &\,-\mathcal{H}\left(2\Psi+\int_r\,(\dot\Phi+\dot\Psi)\right)-\dot\Phi-n^i\p_i\Phi \nonumber \\ 
    &-\frac{1}{r}(\Phi-\Psi)+\frac{1}{r^2}\int_r(\Phi-\Psi) \nonumber \\ 
	&\,+\frac{1}{2r^2}\int_r r'^2\nabla_\perp^2(\Phi+\Psi)+\frac{1}{r^2}\int_r r'(\dot\Phi+\dot\Psi) \nonumber \\
	&\,+\frac{2}{r^2}\int_r\Psi+\frac{1}{r^2}\int_r\int_{r'}\,(\dot\Phi+\dot\Psi)\Bigg]\;.
\end{align}
The difference $\delta k^0_o-\delta\omega_o$ can be replaced using the perturbations to the frequency \eqref{eq:domega=dk0+psi-nv} evaluated at the observer 
\begin{equation}
	\delta\omega_o-\delta k^0_o=\Psi_o-n^iv_{oi}\;.
\end{equation}
Here, $v^i=\d x^i/\d\eta$ is the peculiar velocity. With this, we have
\begin{align}
	\delta\tilde\vartheta(\eta) =&\, a^{-2}\Bigg[\left(\mathcal{H}-\frac{1}{r}\right)(\Psi_o+n^iv_{oi}) \nonumber \\ 
    &\,-\mathcal{H}\left(2\Psi+\int_r\,(\dot\Phi+\dot\Psi)\right)-\dot\Phi-n^i\p_i\Phi \nonumber \\
	&\,-\frac{1}{r}(\Phi-\Psi)+\frac{1}{r^2}\int_r(\Phi-\Psi) \nonumber \\ 
	&\,+\frac{1}{2r^2}\int_r r'^2\nabla_\perp^2(\Phi+\Psi)+\frac{1}{r^2}\int_r r'(\dot\Phi+\dot\Psi) \nonumber \\
	&\,+\frac{2}{r^2}\int_r\Psi+\frac{1}{r^2}\int_r\int_{r'}\,(\dot\Phi+\dot\Psi)\Bigg]\;.
\end{align}
We are now almost done, but the expression is still a function of the background conformal time coordinate $\eta$ and includes the background direction of the wave vector $n^i$, neither of which we have observational access to. Our observations are instead a function of the redshift $\tilde z$ and the observed direction $e^i_o$. Note that the observed direction is the same for $g_{\mu\nu}$ and $\tilde g_{\mu\nu}$ due to its definition $e^i=-\hat k^i/\hat k^0$. We therefore omit the tilde. $n^i$ is easy to replace because the background is Minkowski space and therefore
\begin{equation}
    \label{eq:ei=-ni}
    e^i =-n^i + 1\textrm{st order terms}\,.
\end{equation}
Since $n^i$ only appears multiplied by terms that are already 1st order, we can make the substitution $n^i=-e^i$. Since $n^i$ is constant along the null geodesic we can pick $e^i$ at any point we like and choose $e^i_o$. In order to replace the background conformal time by the observed redshift we use the following Taylor expansion
\begin{align}
	f(\eta(\tilde{\bar z}))\equiv f(\tilde{\bar z}) &= f(\tilde z-\delta \tilde z) \nonumber \\ 
    &\approx f(\tilde z)-\frac{\d f}{\d \tilde{\bar z}}\delta \tilde z \nonumber \\ 
    &=f(\tilde z)+\frac{\d f}{\d\eta}\mathcal{H}^{-1}\delta z\;,
\end{align}
where the perturbations to the redshift are given by \eqref{eq:pertb_z}
\begin{align}
	\delta z &=-n^i[ v_i]^{\eta}_{\eta_o}-[\Psi]_{\eta_o}^{\eta}+\int_\eta\,(\dot\Phi+\dot\Psi)\;.
\end{align}
At first order in perturbation theory, these corrections only have to be applied to background terms. From \eqref{eq:theta_rw} with $K=0$, we find the derivative 
\begin{align}
	\frac{\d \tilde{\bar\vartheta}}{\d\eta}&=a^{-2}\mathcal{H}\left(-2\mathcal{H}+\frac{2}{r}+\frac{\dot{\mathcal{H}}}{\mathcal{H}}-\frac{1}{\mathcal{H}r^2}\right)\;,
\end{align}
Combining everything, we finally obtain 
\begin{align}
	\label{eq:theta_final_result}
	\delta\tilde\vartheta(\tilde z) =&\,(1+\tilde z)^2\Bigg[\left(-\mathcal{H}+\frac{1}{r}+\frac{\dot{\mathcal{H}}}{\mathcal{H}}-\frac{1}{\mathcal{H}r^2}\right)(\Psi_o-e^i_o v_{oi}) \nonumber \\ 
	&\,+\left(-2\mathcal{H}+\frac{2}{r}+\frac{\dot{\mathcal{H}}}{\mathcal{H}}-\frac{1}{\mathcal{H}r^2}\right)e^i_o v_i \nonumber \\
	&\,-\left(\frac{2}{r}+\frac{\dot{\mathcal{H}}}{\mathcal{H}}-\frac{1}{\mathcal{H}r^2}\right)\Psi \nonumber \\
    &\,-\left(-\mathcal{H}+\frac{2}{r}+\frac{\dot{\mathcal{H}}}{\mathcal{H}}-\frac{1}{\mathcal{H}r^2}\right)\int_r\,(\dot\Phi+\dot\Psi) \nonumber \\
	&\,-\dot\Phi+e^i_o\p_i\Phi-\frac{1}{r}(\Phi-\Psi)+\frac{1}{r^2}\int_r(\Phi-\Psi) \nonumber \\ 
	&\,+\frac{1}{2r^2}\int_r r'^2\nabla_\perp^2(\Phi+\Psi)+\frac{1}{r^2}\int_r r'(\dot\Phi+\dot\Psi) \nonumber \\
	&\,+\frac{2}{r^2}\int_r\Psi+\frac{1}{r^2}\int_r\int_{r'}\,(\dot\Phi+\dot\Psi)\Bigg]\;,
\end{align}
where $\mathcal{H}$ and $r$ are now to be understood as the usual background functions evaluated at the observed redshift $\tilde z$. Now that we are finished with the derivation we will drop the tildes and it is assumed that any result is given in the expanding spacetime. 
\\ \\ 
We have derived this result using the perpendicular Laplacian $\nabla_\perp^2$ in order to more easily compare the power spectra derived in the next section to those describing the fluctuations of the angular diameter distance which are most naturally expressed using $\nabla_\perp^2$. For $\vartheta$, we will see that on very large scales the term proportional to $e^i_o\p_i$ and the term proportional to $\nabla_\perp^2$ have similar magnitude. Since they are closely related, it can thus be useful to express the result for $\vartheta$ simply in terms of $\nabla^2$. A short calculation (see Appx.~\ref{appx:rewriting_theta}) shows 
\begin{align}
    \label{eq:theta_final_v2}
	\delta\vartheta(z) =&\,(1+z)^2\Bigg[\left(-\mathcal{H}+\frac{1}{r}+\frac{\dot{\mathcal{H}}}{\mathcal{H}}-\frac{1}{\mathcal{H}r^2}\right)(\Psi_o-e^i_o v_{oi}) \nonumber \\ 
	&\,+\left(-2\mathcal{H}+\frac{2}{r}+\frac{\dot{\mathcal{H}}}{\mathcal{H}}-\frac{1}{\mathcal{H}r^2}\right)e^i_o v_i \nonumber \\
	&\,-\left(\frac{2}{r}+\frac{\dot{\mathcal{H}}}{\mathcal{H}}-\frac{1}{\mathcal{H}r^2}\right)\Psi \nonumber \\
    &\,-\left(-\mathcal{H}+\frac{2}{r}+\frac{\dot{\mathcal{H}}}{\mathcal{H}}-\frac{1}{\mathcal{H}r^2}\right)\int_r\,(\dot\Phi+\dot\Psi) \nonumber \\
	&\,-\frac{1}{2}(3\dot\Phi+\dot\Psi)+\frac{1}{2}e^i_o\p_i(\Phi-\Psi)-\frac{1}{r}(\Phi-\Psi)\nonumber \\ 
    &\,+\frac{1}{r^2}\int_r(\Phi-\Psi)+\frac{1}{2r^2}\int_r r'^2\nabla^2(\Phi+\Psi) \nonumber \\ 
	&\,-\frac{1}{2r^2}\int_r r'^2(\ddot\Phi+\ddot\Psi)+\frac{2}{r^2}\int_r r'(\dot\Phi+\dot\Psi) \nonumber \\
	&\,+\frac{2}{r^2}\int_r\Psi+\frac{1}{r^2}\int_r\int_{r'}\,(\dot\Phi+\dot\Psi)\Bigg]\;.
\end{align}
While this has more terms than the earlier expression, we will later see that it has fewer non-negligible terms.

\section{Power Spectra} 
\label{sect:power_spectra}
The fluctuations to $\vartheta$ are most naturally studied in terms of its angular power spectrum $C_l$. Calculating angular power spectra for long expressions, such as \eqref{eq:theta_final_result} is somewhat tedious but there is a very generic pattern that makes reading off the power spectra almost trivial. We derive the general structure in Appx.~\ref{appx:powerspectra} and give the result here. We assume that each of the random variables $f_i\in\{\Phi, \Psi, v^i\}$ can in Fourier space be expressed through a single initial random variable, which we choose to be the Bardeen potential $\Phi$, according to
\begin{equation}
    f(\b k, z)=T_f(k,z)\Phi_\mathrm{in}\,(\b k)\;.
\end{equation}
$T_f(k, z)$ is a deterministic function that depends only on the absolute magnitude of the wave number and the redshift. $\Phi_\mathrm{in}(\b k)$ carries all the angular dependence and has power spectrum
\begin{equation}
	\langle \Phi_\textrm{in}(\b k)\Phi_\textrm{in}^*(\b k')\rangle = (2\pi)^3 \delta (\b k -\b k')\mathcal{P}(k)\;.
\end{equation}
For any observable on the sphere $O(\b e_o, z)$ which can be written as 
\begin{equation}
    O(\b e_o, z) =\sum_i F_i(\b e_o, z)\;,\;\; \textrm{where}\;\; F_i \in \left\{f\;,\b e_o\cdot\nabla f\;, \nabla^2_\Omega f \right\}\;,
\end{equation} 
the angular power spectrum can then be obtained as
\begin{equation}
    \label{eq:generic_Cl}
    C_l^{O_1 O_2}(z_1, z_2)= \frac{2}{\pi}\int_0^\infty \d k k^2\; \mathcal{P}(k)\mathcal{T}^W_{O_1}(k,z_1,l)\mathcal{T}^W_{O_2}(k,z_2,l)\;,
\end{equation}
where
\begin{equation}
    \mathcal{T}^W_O(k, z, l)=\int_0^\infty\d z'\,W(z,z')\sum_i\mathcal{T}_{F_i}(k,z,l)\;
\end{equation}
and
\begin{equation}
    \label{eq:general_transfer}
    \mathcal{T}_O(k,z,l)=\sum_i\mathcal{T}_{f_i}(k,z,l)\;.
\end{equation}
We have here distinguished between terms $F_i$ with different angular dependence because they generate different contributions to the power spectrum. For the different $F_i$ we find
\begin{align}
    \mathcal{T}_f(k,z,l) &= T_f(k,z)j_l(k,r(z))\, \\
    \mathcal{T}_{\b e_o\cdot\nabla f}(k,z,l) &= kT_f(k,z) j'_l(k,r(z))\, \\ 
    \mathcal{T}_{\nabla^2_\Omega f}(k,z,l) &= -l(l+1)T_f(k,z) j_l(k,r(z))\;, 
\end{align}
where the $j_l$ denote the spherical Bessel functions of the first kind and the $j_l'$ are the first derivative w.r.t. to their argument. 

In the general formula \eqref{eq:generic_Cl} for the angular power spectrum we have allowed for two different observables $O_1, O_2$ at two different redshifts, to allow the calculation of cross-spectra. In the following, we will only consider auto-correlations. The window function $W(z, z')$ allows us to take into account that observations are typically made in redshift bins with particular redshift distributions. We will consider a Dirac delta function as well as a Gaussian distribution of redshifts with mean $z$ and width $\sigma$ as window functions in the following.
\\ \\
With the general formalism ready, we can now read off the transfer function \eqref{eq:general_transfer} for $\delta\vartheta$. Writing the velocity as the gradient of a potential such that $e^i_o v_i=-\b e_o\cdot\nabla\,v$, and using that $\nabla^2_\perp=1/r^2\nabla^2_\Omega$ we find from \eqref{eq:theta_final_result} that
\begin{align}
	\mathcal{T}_{\delta\vartheta}&\,(k,z,l) = (1+z)^2\Bigg[\nonumber \\
    &\,-\left(-2\mathcal{H}+\frac{2}{r}+\frac{\dot{\mathcal{H}}}{\mathcal{H}}-\frac{1}{\mathcal{H}r^2}\right)kj'_l(kr)T_v \nonumber\\ 
	&\,-\left(\frac{2}{r}+\frac{\dot{\mathcal{H}}}{\mathcal{H}}-\frac{1}{\mathcal{H}r^2}\right)j_l(kr)T_\Psi \nonumber \\
	&\,-\left(-\mathcal{H}+\frac{2}{r}+\frac{\dot{\mathcal{H}}}{\mathcal{H}}-\frac{1}{\mathcal{H}r^2}\right)\int_{r'}j_l(kr')(T_{\dot\Phi}+T_{\dot\Psi})\nonumber \\
	&\,-j_l(kr)T_{\dot\Phi}+kj'_l(kr)T_\Phi-\frac{1}{r}(T_\Phi-T_\Psi) \nonumber \\ 
    &\,+\frac{1}{r}\int_{r'} (T_\Phi-T_\Psi)\nonumber \\ 
	&\,-l(l+1)\frac{1}{2r^2} \int_{r'} j_l(kr')(T_\Phi+T_\Psi) \nonumber \\ 
    &\,+\frac{1}{r^2}\int_{r'} r'j_l(kr')(T_{\dot\Phi}+T_{\dot\Psi}) \nonumber \\
	&\,+\frac{2}{r^2} \int_{r'}j_l(kr')T_\Psi+\frac{1}{r^2} \int_{r'}\int_{r''}j_l(kr'')(T_{\dot\Phi}+T_{\dot\Psi})\Bigg]\;.
\end{align}
Here we have neglected the observer terms $\Psi_o$ and $v^i_o$. Our own peculiar motion introduces a dipole, that we study separately in Sect.~\ref{sect:dipole}.

For the alternative expression \eqref{eq:theta_final_v2} we identify the transfer function 
\begin{align}
	\mathcal{T}_{\delta\vartheta}&\,(k,z,l) = (1+z)^2\Bigg[\nonumber \\
    &\,-\left(-2\mathcal{H}+\frac{2}{r}+\frac{\dot{\mathcal{H}}}{\mathcal{H}}-\frac{1}{\mathcal{H}r^2}\right)kj'_l(kr)T_v \nonumber\\ 
	&\,-\left(\frac{2}{r}+\frac{\dot{\mathcal{H}}}{\mathcal{H}}-\frac{1}{\mathcal{H}r^2}\right)j_l(kr)T_\Psi \nonumber \\
	&\,-\left(-\mathcal{H}+\frac{2}{r}+\frac{\dot{\mathcal{H}}}{\mathcal{H}}-\frac{1}{\mathcal{H}r^2}\right)\int_{r'}j_l(kr')(T_{\dot\Phi}+T_{\dot\Psi})\nonumber \\
	&\,-\frac{1}{2}j_l(kr)(3T_{\dot\Phi}+T_{\dot\Psi})+\frac{1}{2}kj'_l(kr)(T_\Phi-T_\Psi) \nonumber \\ 
    &\,-\frac{1}{r}(T_\Phi-T_\Psi)+\frac{1}{r}\int_{r'} (T_\Phi-T_\Psi) \nonumber \\ 
	&\,-\frac{1}{2r^2} \int_{r'} r'^2 j_l(kr')\left[k^2(T_\Phi+T_\Psi)+(T_{\ddot\Phi}+T_{\ddot\Psi}) \right]  \nonumber \\ 
    &\,+\frac{2}{r^2}\int_{r'} r'j_l(kr')(T_{\dot\Phi}+T_{\dot\Psi})\nonumber \\ 
	&\,+\frac{2}{r^2} \int_{r'}j_l(kr')T_\Psi+\frac{1}{r^2} \int_{r'}\int_{r''}j_l(kr'')(T_{\dot\Phi}+T_{\dot\Psi})\Bigg]\;, 
\end{align}
where we used that $\nabla^2=-k^2$ in Fourier space. 
\\ \\ 
As $\vartheta$ is constructed from the angular diameter distance, we will compare our results to the fluctuations in the angular diameter distance in order to understand if we gain anything by constructing $\vartheta$ rather then directly studying the fluctuations to $D_A$. A similar exercise to that which lead to \eqref{eq:theta_final_result} \cite{Oestreicher_2026_PhD} (see also \cite{Bonvin_2006_Fluctuations_Luminosity_Distance, Bonvin_2008_pec_motion_weak_lensing, Umeh_2014_DA_second_order_derivation})  yields 
\begin{align}\label{eq:da}
	\frac{\delta D_A}{\bar D_A}(z)=&\,\frac{1}{\mathcal{H} r}v^i_o e_{oi}+\left(1-\frac{1}{\mathcal{H} r}\right)\Psi_o \nonumber \\ 
    &\,-\Phi +\left(1-\frac{1}{\mathcal{H} r}\right)\left(v^i e_{oi}-\Psi-\int_r(\dot\Phi+\dot\Psi)\right)\nonumber \\
	&\,+\frac{1}{r}\int_r(\Phi+\Psi)+\frac{1}{2r}\int_r r'(r'-r)\nabla_\perp^2(\Phi+\Psi)\;.
\end{align}
As before, writing $e^i_o v_i=-\b e_o\cdot\nabla\,v$ and using $\nabla^2_\perp=1/r^2\nabla^2_\Omega$, we read off the transfer function
\begin{align}
	\mathcal{T}_{\delta D_A/\bar D_A}&\,(k,z,l)=-j_l(kr)T_\Phi \nonumber \\ 
    &\,-\left(1-\frac{1}{\mathcal{H} r}\right)\left(k j'_l(kr) T_v+j_l(kr) T_\Psi\right) \nonumber \\
	&\,-\left(1-\frac{1}{\mathcal{H} r}\right)\int_r j_l(kr')(T_{\dot\Phi}+T_{\dot\Psi})\nonumber \\
	&\,+\frac{1}{ r}\int_rj_l(kr')(T_\Phi+T_\Psi) \nonumber \\ 
	&\,-l(l+1)\frac{1}{2r}\int_r \frac{(r'-r)}{r'}j_l(kr')(T_\Phi+T_\Psi)\;.
\end{align}
\\ \\ 
Until now, we have not assumed any particular theory of gravity and the results hold in general, for any FLRW spacetime with scalar fluctuations. In order to show some numeric results we now evaluate the result using the field equations of general relativity. To first order in perturbation theory for the metric with line element \eqref{eq:metric} Einstein's field equations read \cite{Baumann_2022_Cosmology}
\begin{align}
    \nabla^2\Phi-3\mathcal{H}(\dot\Phi+\mathcal{H}\Psi)&=4\pi G_N a^2\delta\rho\;, \\
    \dot\Phi+\mathcal{H}\Psi &= 4\pi G_N a^2 (\bar\rho+\bar p)v\;, \\
    \Phi-\Psi&=8\pi G_N a^2 \Pi\;,  \\
    \ddot\Phi+\mathcal{H}\dot\Psi+2\mathcal{H}\dot\Phi+\frac{1}{3}\nabla^2(\Psi&-\Phi)+(2\dot{\mathcal{H}}+\mathcal{H}^2)\Psi \nonumber \\ &= 4\pi G_N a^2\delta p \;.
\end{align}
Here $\bar\rho$ and $\bar p$ are the mean density and pressure, $\delta\rho$ and $\delta p$ describe their fluctuations, $v$ is the velocity potential defined via $v_i = -\nabla_i v$, and $\Pi$ is the scalar anisotropic stress. If the universe is dominated by pressureless dark matter and anisotropic stress is negligible, the later three imply that 
\begin{align}
	T_\Phi&=T_\Psi \\
	T_v&=\frac{2}{3}(H_0^2\Omega_{m0}(1+z))^{-1}g(T_\Phi'+\mathcal{H}T_\Psi) \\
	T_\Phi(k,z)&=g(z)T_\Phi(k,z=0)\;,
\end{align}
where the growth factor $g(z)$ obeys
\begin{align}
    \ddot g+3\mathcal{H}\dot g+(2\dot{\mathcal{H}}+\mathcal{H}^2)g=0
\end{align}
and we introduced the current day matter density parameter in the usual manner, i.e. $\Omega_{m0}=4\pi G_N\bar\rho a^{3}/(3H_0^2)$. The growth factor $g$ is related to the usual growth factor of the density perturbations $D_+$ as 
$g=D_+/a$. We obtain $T_\Phi(k,z=0)$ and $D_+(z)$ from \texttt{CLASS}\footnote{\url{http://class-code.net}}\ \cite{CLASS_II}. We assume a standard power law for the primordial power spectrum, i.e.
\begin{equation}
    \mathcal{P}(k)=\frac{2\pi^2}{k^3}A_s\left(\frac{k}{k^*}\right)^{n_s-1}\;,
\end{equation}
with $n_s$ and $A_s$ specified in the next section. We now have everything necessary to calculate the power spectra numerically. In order to solve the highly oscillating integrals over the appearing spherical Bessel functions, we use the library \texttt{FFTlog-and-beyond}\footnote{\url{https://github.com/xfangcosmo/FFTLog-and-beyond}}\cite{Fang_2020_FFTlog}. The details of our numerical implementation are given in Appx.~\ref{appx:power_spectra_numerics}.

\section{Numerical Results}
\label{sect:numerical_results}
In this section we show some numeric results for the observer induced dipole and the power spectra. We will calculate power spectra both using the linear theory presented above and from ray tracing through a simulation made with \texttt{gevolution}. Unless otherwise specified we use the default cosmological parameters of \texttt{gevolution} which are: $A_s = 2.215\times 10^{-9}$, $n_s = 0.9619$, $k_{\rm pivot} = 0.05$, $h = 0.67556$, $h^2\Omega_{\rm b} = 0.022032$, $h^2\Omega_{\rm cdm} = 0.12038$ and $N_{eff} = 3.046$.

\begin{figure}
    \centering
    \includegraphics[width=0.8\linewidth]{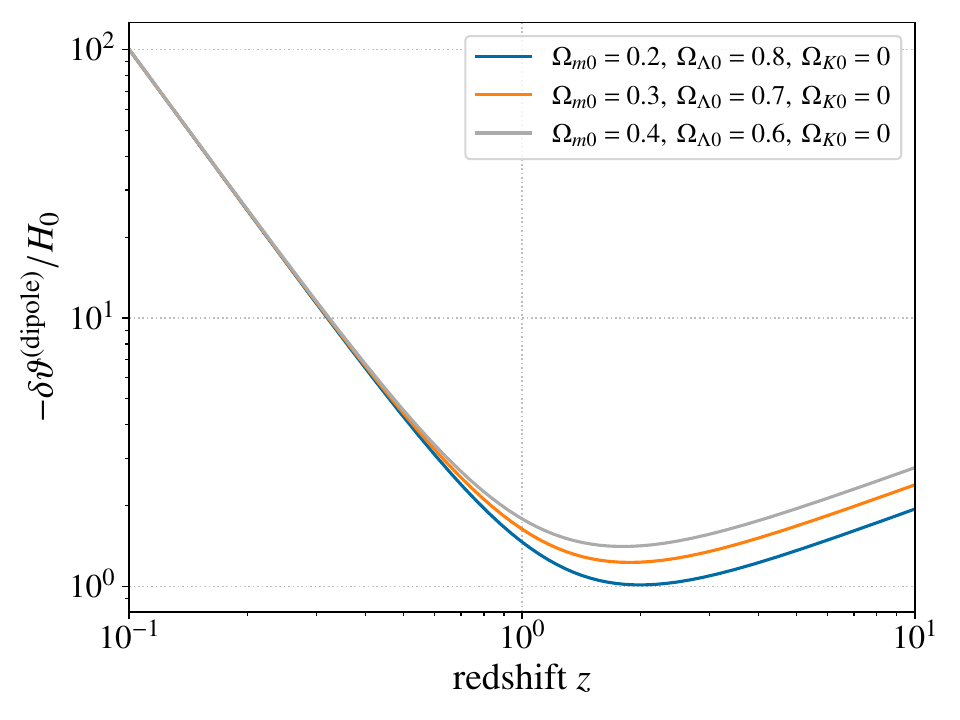}
    \includegraphics[width=0.8\linewidth]{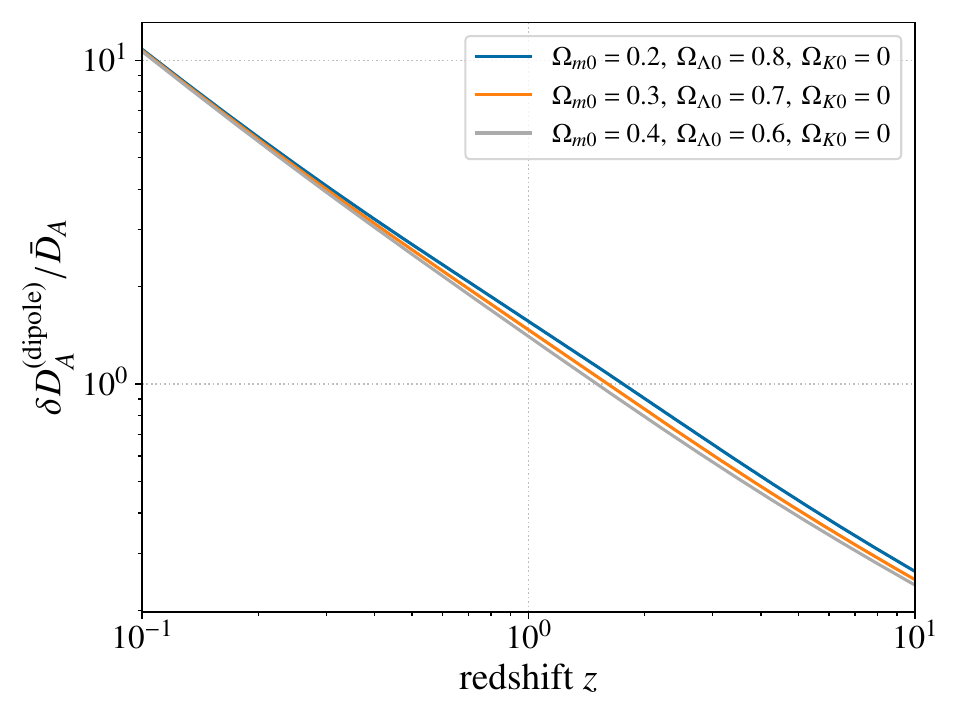}
    \caption{Redshift dependence of the dipole induced by our peculiar motion w.r.t to the cosmic rest frame for $-\vartheta/H_0$ and $-\delta D_A/\bar D_A$ for three $\Lambda$CDM models. The model parameters are indicated in the legends.}
    \label{fig:dipole}
\end{figure}

\subsection{Dipole}
\label{sect:dipole}
We will first briefly consider the dipole introduced to $\vartheta$ and $\delta D_A/\bar D_A$ by our own motion with respect to the CMB, i.e. the $e^i_o v_{oi}$ term. Assuming that this term dominates the dipole, we can see from \eqref{eq:theta_final_result} and \eqref{eq:da} that the dipoles should have redshift dependence
\begin{align}\label{eq:theta_dipole}
 	\delta\tilde\vartheta^{(\rm dipole)}(z) = (1+z)^2\left(-\mathcal{H}+\frac{1}{r}+\frac{\dot{\mathcal{H}}}{\mathcal{H}}-\frac{1}{\mathcal{H}r^2}\right)
 \end{align}
and
\begin{align}\label{eq:da_dipole}
    \frac{\delta D_A}{\bar D_A}^{(\rm dipole)}(z)=&\,\frac{1}{\mathcal{H} r}.
\end{align}
We visualize this redshift dependence in Fig.~\ref{fig:dipole} for a perturbed spacetime based on the same flat $\Lambda$CDM models considered in Fig.~\ref{fig:monopole}.  While the dipole does not provide independent information beyond that contained in $H$ and $D_A$ individually, its characteristic redshift dependence offers a cross-check of the consistency between the quantities by verifying that the inferred expansion history and redshift-distance relation satisfy the relations expected by an assumed cosmology. In particular, a measured dipole inconsistent with the prediction would indicate problems in one of the datasets, systematics, $D_A'$, or a breakdown of the assumptions entering the derivation. 

\begin{figure*}[ht]
    \centering
    \includegraphics[width=0.4\linewidth]{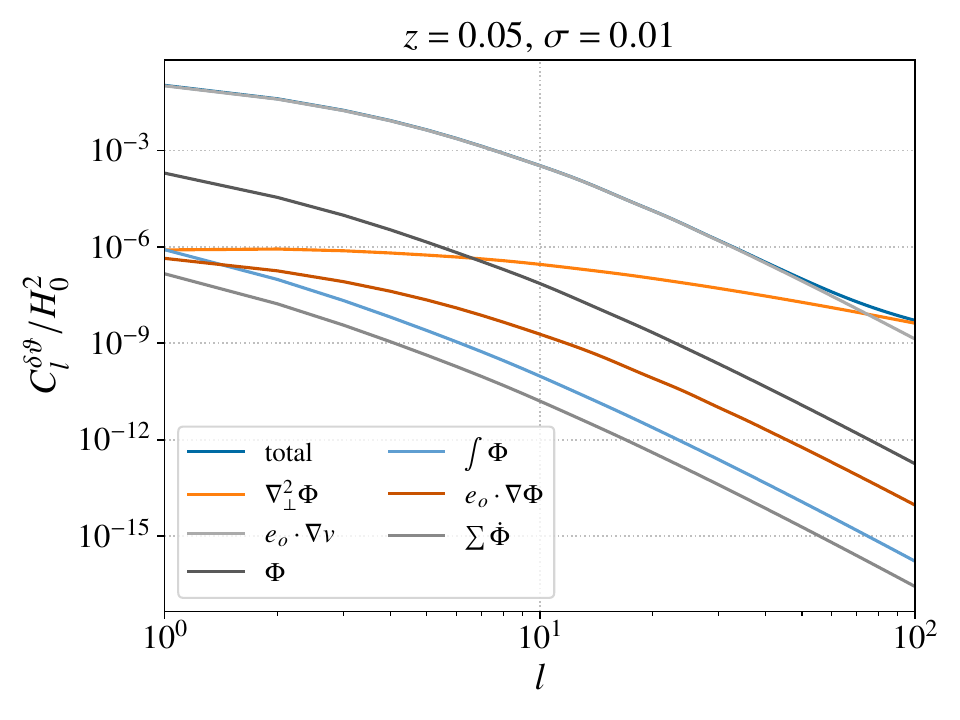}%
    \includegraphics[width=0.4\linewidth]{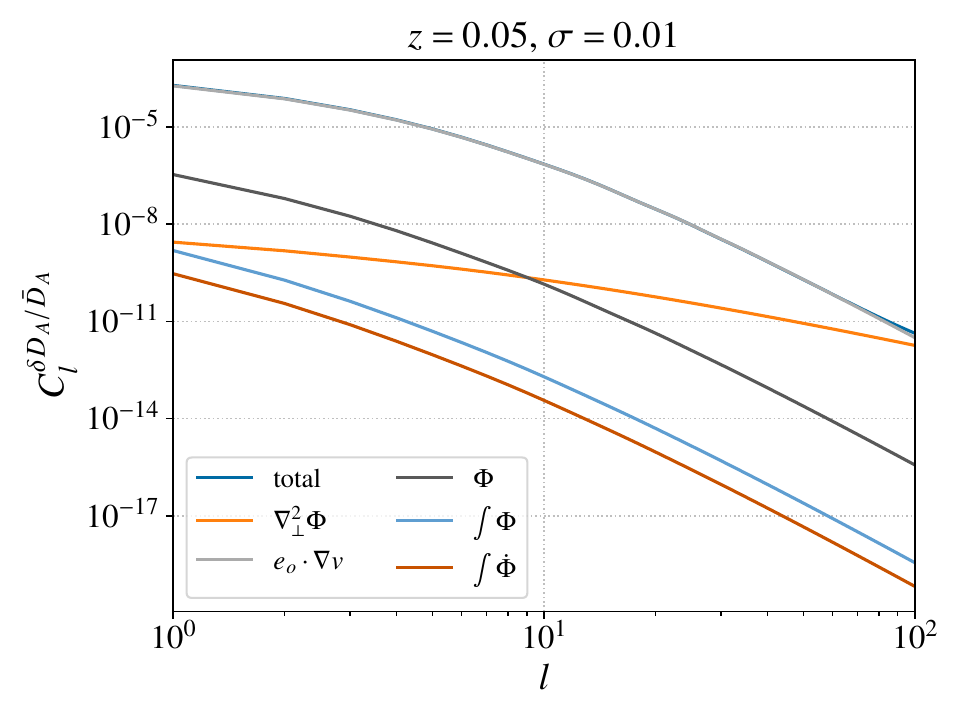}
    \includegraphics[width=0.4\linewidth]{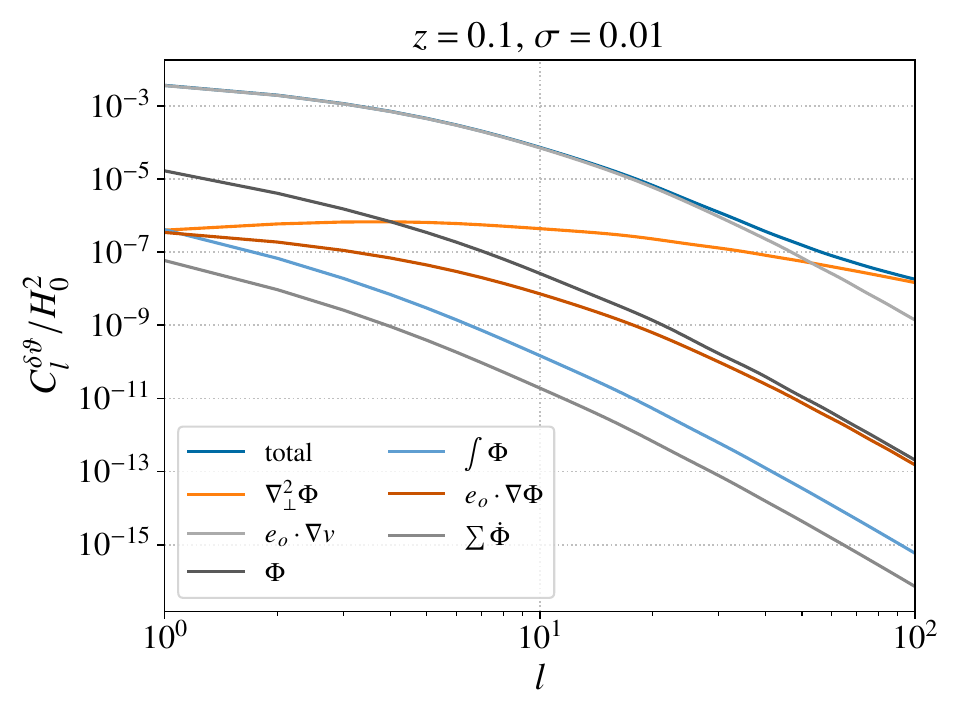}%
    \includegraphics[width=0.4\linewidth]{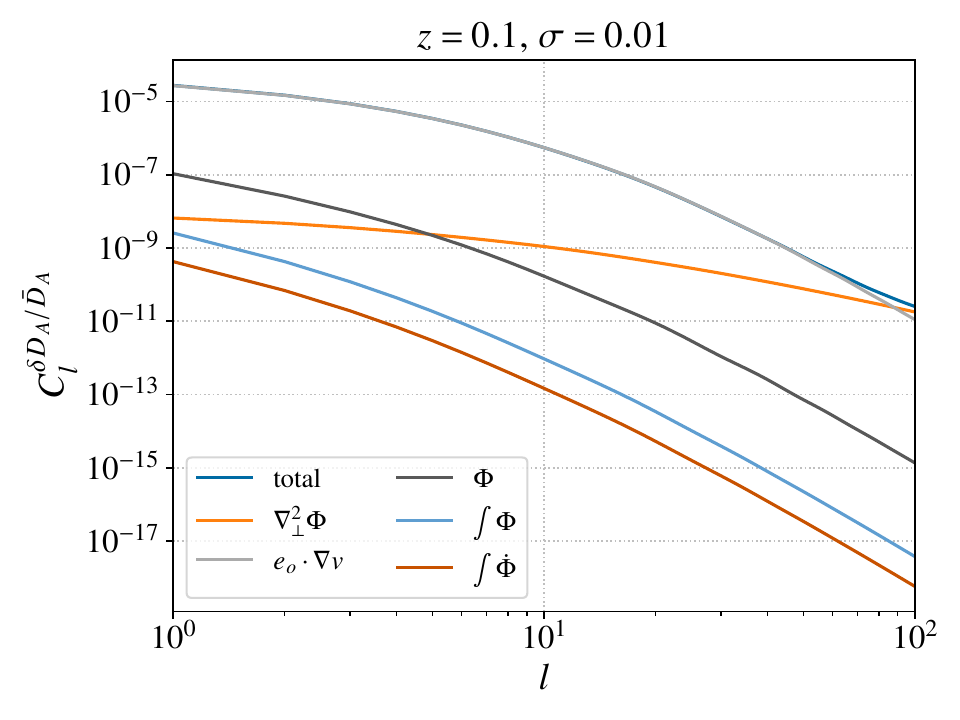}
    \caption{Angular power spectra $C_l$ for the fluctuations to the optical expansion scalar $\delta\vartheta$ \eqref{eq:theta_final_result} (left) and the fluctuations to the angular diameter distance $\delta D_A/\bar D_A$ \eqref{eq:da} (right). Evaluated for a Gaussian window function, with mean $z=0.05,0.1$ and width $\sigma=0.01$. We show the individual contributions to the power spectra as well as the total spectrum, for $\delta\vartheta$ we combine all terms that contain time derivatives of the Bardeen potential $\Phi$.}
    \label{fig:cl_gauss_lowz}
\end{figure*}

\begin{figure*}[ht]
    \centering
    \includegraphics[width=0.4\linewidth]{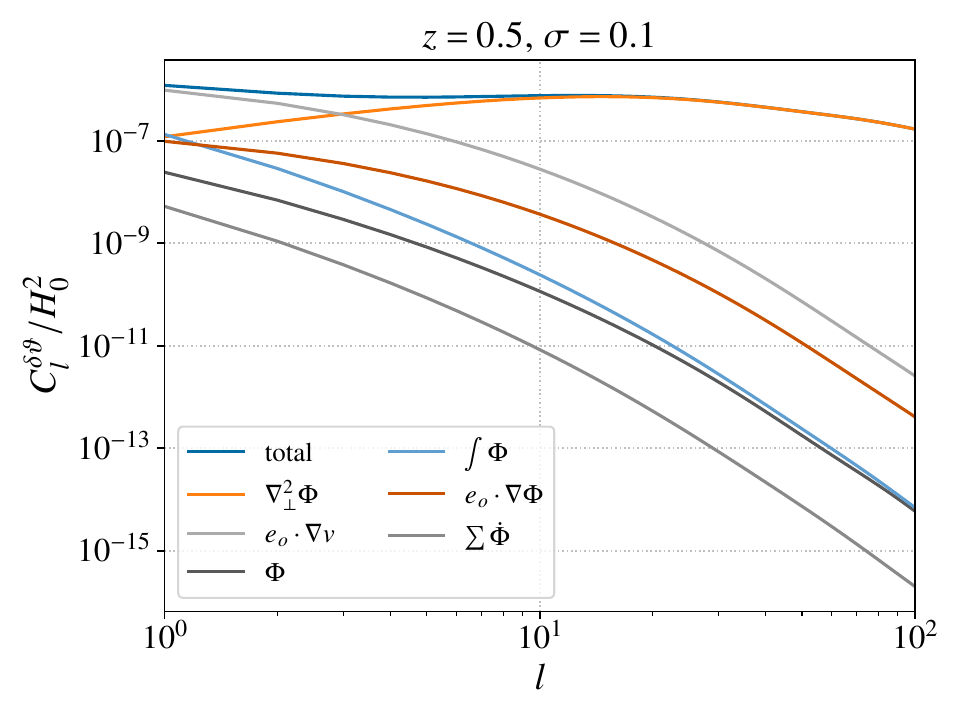}%
    \includegraphics[width=0.4\linewidth]{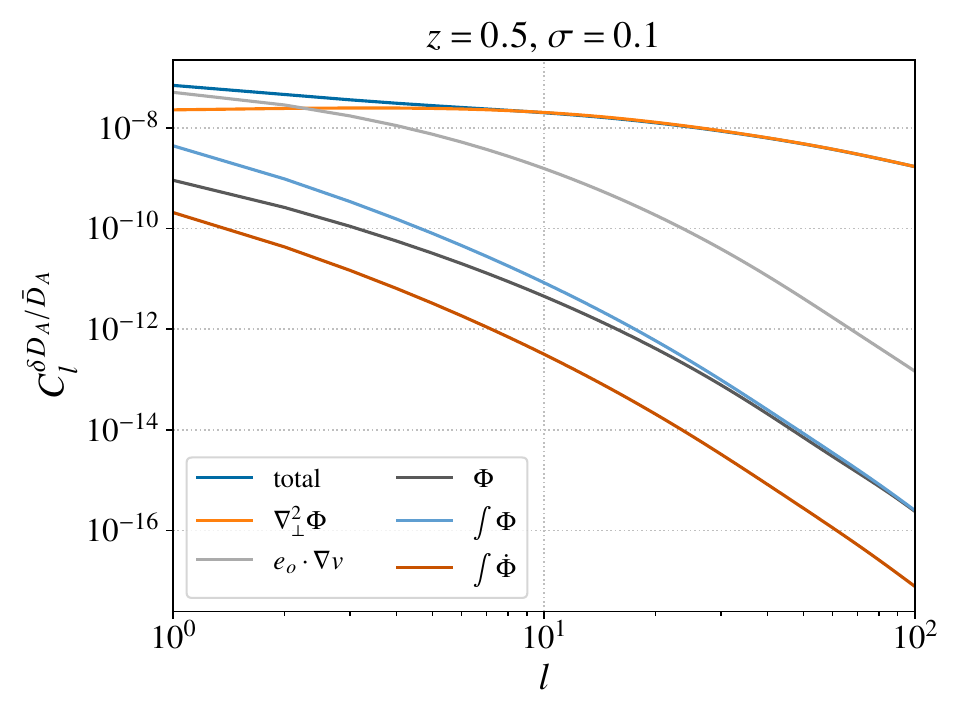}
    \includegraphics[width=0.4\linewidth]{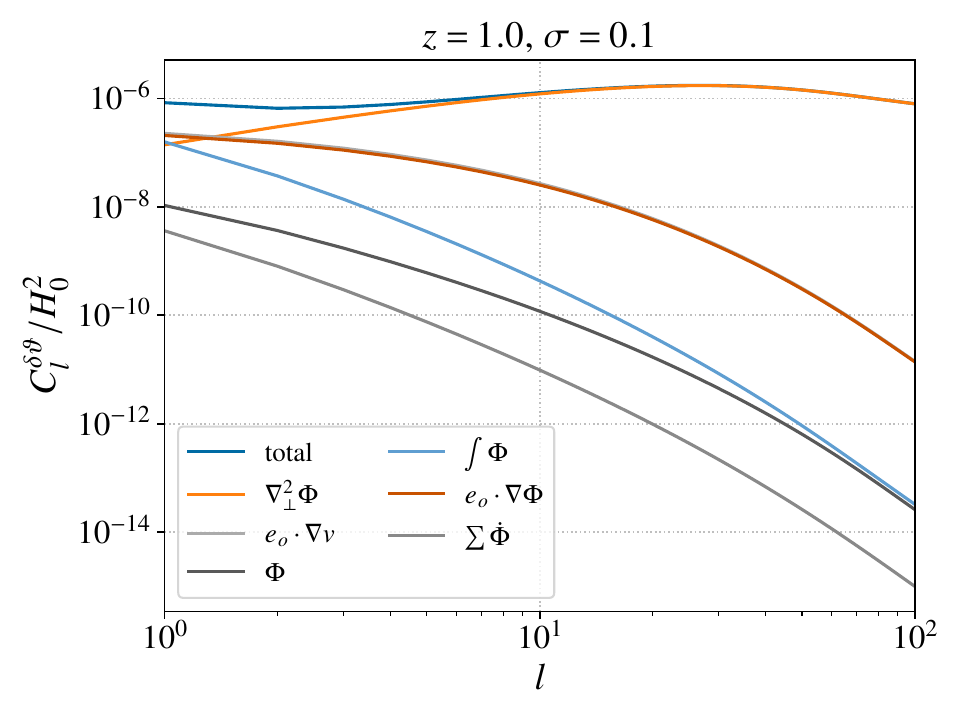}%
    \includegraphics[width=0.4\linewidth]{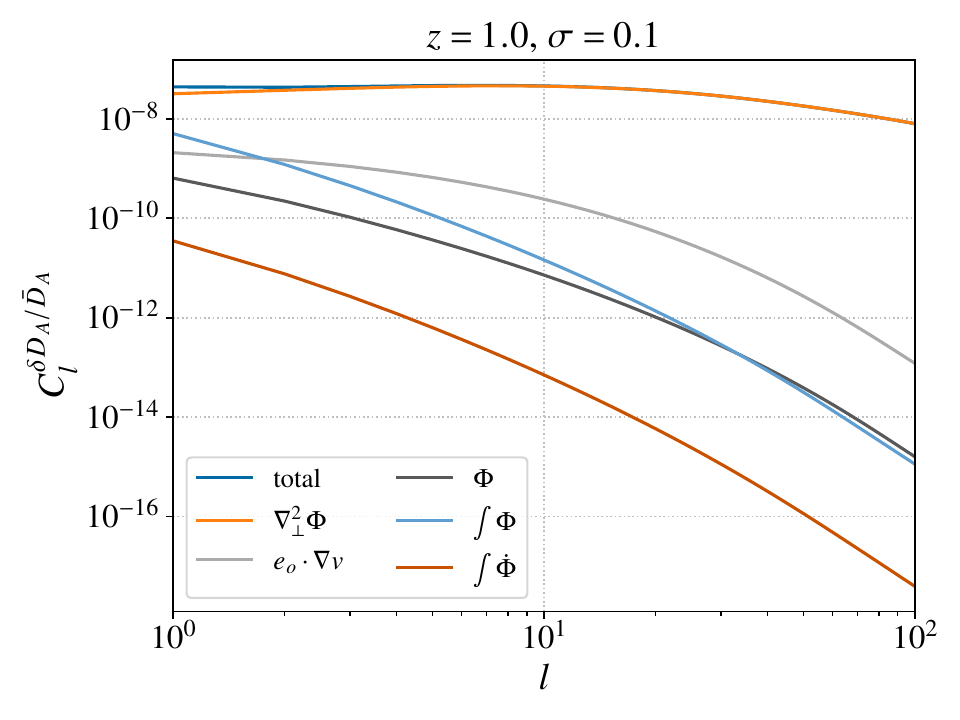}
    \includegraphics[width=0.4\linewidth]{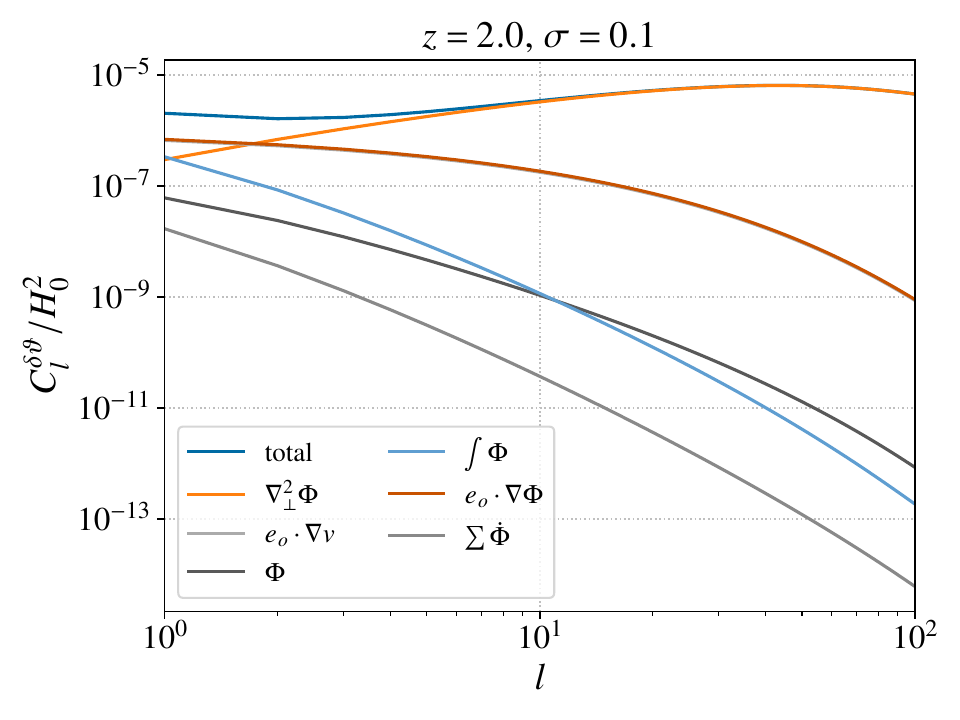}%
    \includegraphics[width=0.4\linewidth]{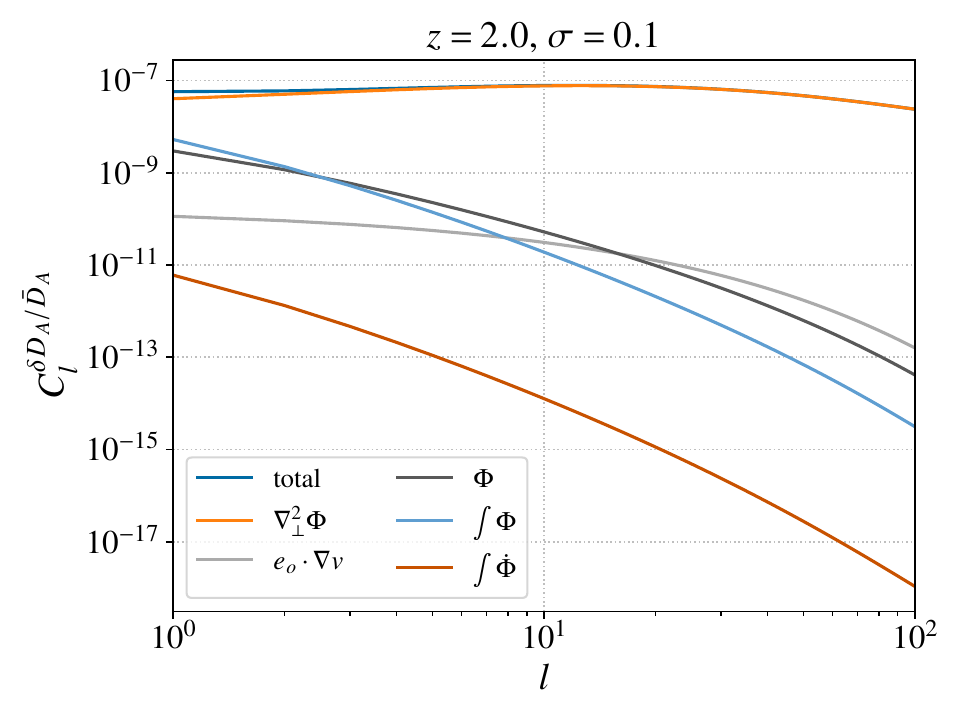}
    \caption{Same as Fig.~\ref{fig:cl_gauss_lowz}, but for Gaussian windows with mean $z=0.5,1.0,2.0$ and width $\sigma=0.1$.}
    \label{fig:cl_gauss_highz}
\end{figure*}

\subsection{Perturbation Theory}
We show the linear power spectra of $\delta\vartheta$ and $\delta D_A$ as given by \eqref{eq:theta_final_result} and \eqref{eq:da} for a range of different redshifts in Fig.~\ref{fig:cl_gauss_lowz} and \ref{fig:cl_gauss_highz}, with the power spectra split into their various contributing terms. Note that cross terms also appear in the total power spectrum but that their individual contributions are not shown. 

Fig.~\ref{fig:cl_gauss_lowz} shows the linear power spectra for a Gaussian window function with means $z=0.05$ and $0.1$ and width $\sigma=0.01$. For both the fluctuations in $\vartheta$ and $D_A$ the term arising from the peculiar velocity by far dominates the total power spectra, and all other terms can be neglected until around $l \approx 50$, where the contribution arising from the perpendicular Laplacian, often referred to as the lensing potential and closely related to the density \cite{Bonvin_2008_pec_motion_weak_lensing}, starts to dominate the total spectra instead. 

Fig.~\ref{fig:cl_gauss_highz} shows the same spectra but for the  higher redshifts $z=0.5,1.0,2.0$, now with $\sigma=0.1$. For all three redshifts the contribution from the $\nabla_\perp^2\Phi$ term dominates at $l\gtrsim 10$. For $\delta\vartheta$ the contributions from the peculiar velocity, spatial gradient of the potential $e^i_o\nabla_i\Phi$ and the term containing an integral over $\Phi$ contribute significantly at low $l$. For the fluctuations in $D_A$, the term containing the peculiar velocity contributes at $z=0.5$ but quickly becomes entirely negligible for $ z = 1,2$. At these higher redshifts, the fluctuations in $D_A$ are almost perfectly proportional to only the $\nabla_\perp^2\Phi$ term, with only a small contribution from the potential and integrated potential term at very low $l$. This is as expected based on e.g. \cite{Bonvin_2008_pec_motion_weak_lensing, Bonvin_2006_Fluctuations_Luminosity_Distance}.

\begin{figure*}[ht]
    \centering
    \includegraphics[width=0.4\linewidth]{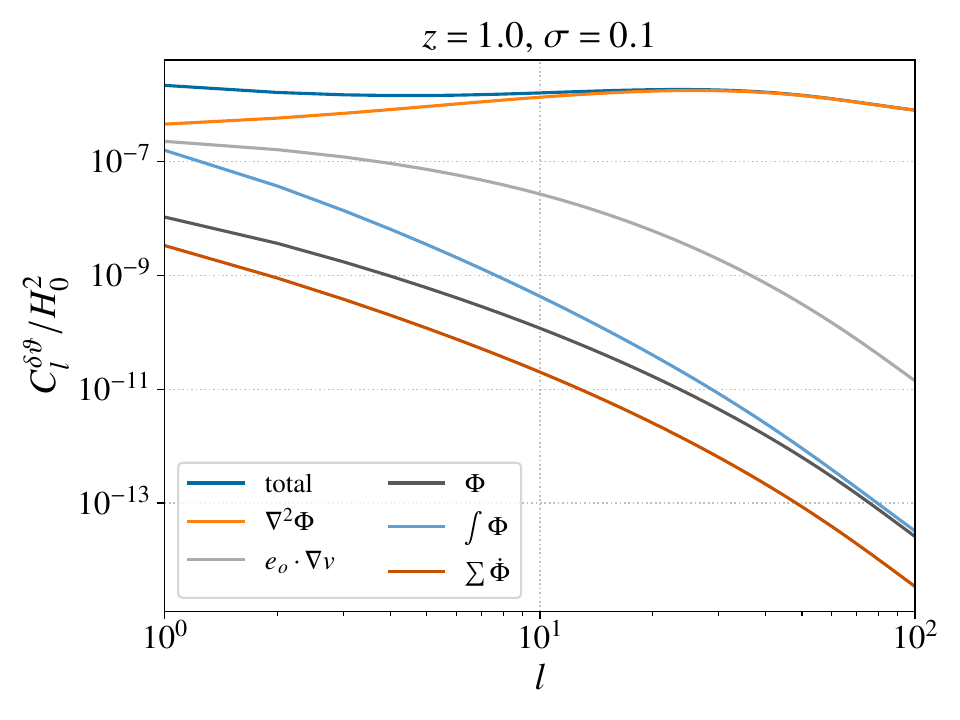}%
    \includegraphics[width=0.4\linewidth]{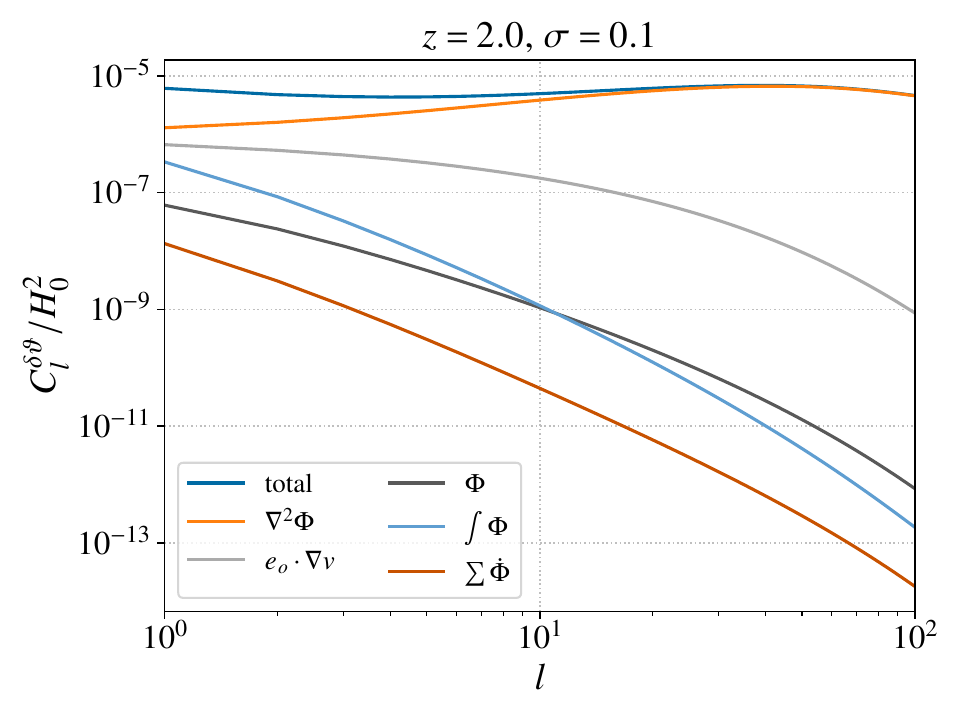}
    \caption{Angular power spectra $C_l$ for the fluctuations to the optical expansion scalar $\vartheta$ according to the expression \eqref{eq:theta_final_v2}. For Gaussian windows with mean $z=1.0,2.0$ and width $\sigma=0.1$. We show the individual contributions to the power spectra as well as the total spectrum, we combine all terms that contain time derivatives of the Bardeen potential $\Phi$.}
    \label{fig:cl_theta_v2}
\end{figure*}

As we have shown in \eqref{eq:theta_final_v2}  the result for $\delta\vartheta$ can be rewritten such that it contains simply the Laplacian $\nabla^2\Phi$ and not the perpendicular Laplacian $\nabla_\perp^2\Phi$. As long as there is no anisotropic stress, the resulting power spectrum for $\delta\vartheta$ then has fewer significant terms. In particular, the term $e^i_o\nabla_i\Phi$ relevant at low $l$ no longer appears. We show the power spectra for this version of $\delta\vartheta$ in Fig.~\ref{fig:cl_theta_v2} for a Gaussian window with mean redshift $z=1.0,2.0$ and width $\sigma=0.1$. 

We now summarize the results by comparing the expressions for $\vartheta$ and $D_A$, considering only the dominant terms. For $\Phi=\Psi$ and neglecting all terms containing a time derivative (since these are subdominant) the expressions \eqref{eq:theta_final_v2} and \eqref{eq:da} for the fluctuations in $\vartheta$ and $D_A$ reduce to 
\begin{align}
    \label{eq:theta_dominant_terms}
	\delta\vartheta(z) \simeq&\,(1+z)^2\Bigg[\left(-\mathcal{H}+\frac{1}{r}+\frac{\dot{\mathcal{H}}}{\mathcal{H}}-\frac{1}{\mathcal{H}r^2}\right)(\Phi_o-e^i_o v_{oi}) \nonumber \\ 
	&\,+\left(-2\mathcal{H}+\frac{2}{r}+\frac{\dot{\mathcal{H}}}{\mathcal{H}}-\frac{1}{\mathcal{H}r^2}\right)e^i_o v_i +\frac{2}{r^2}\int_r\Phi\nonumber \\
	&\,-\left(\frac{2}{r}+\frac{\dot{\mathcal{H}}}{\mathcal{H}}-\frac{1}{\mathcal{H}r^2}\right)\Phi +\frac{1}{r^2}\int_r r'^2\nabla^2\Phi\Bigg]\;.
\end{align}
and 
\begin{align}
    \label{eq:DA_dominant_terms}
	\frac{\delta D_A}{\bar D_A}(z)\simeq&\,\frac{1}{\mathcal{H} r}v^i_o e_{oi}+\left(1-\frac{1}{\mathcal{H} r}\right)\Phi_o\nonumber \\ 
    &\,+\left(1-\frac{1}{\mathcal{H} r}\right)v^i e_{oi}-\left(2-\frac{1}{\mathcal{H} r}\right)\Phi\nonumber \\
	&\,+\frac{2}{ r}\int_r\Phi+\frac{1}{r}\int_r r'(r'-r)\nabla_\perp^2\Phi\;.
\end{align}
The two expressions contain similar terms, but with a different redshift dependence. 
For $z\lesssim0.5$ the power spectra for both these expressions are dominated by the peculiar velocity. For higher redshifts, the fluctuations to $\vartheta$ are dominated by $\nabla^2\Phi$, while those to $D_A$ are dominated by $\nabla_\perp^2\Phi$, especially for large $l$. For high redshift and low $l$, both power spectra have somewhat significant contributions from the peculiar velocity, the potential, and a term integrating the potential along the light path. For $\delta\vartheta$ these contributions are notably more significant, hinting that $\delta\vartheta$ might allow for better constrains of these contributions on large scales, than simply considering $\delta D_A$.  

\subsection{N-body Simulation Results}
\begin{figure*}[ht]
    \centering
    \includegraphics[width=0.4\linewidth]{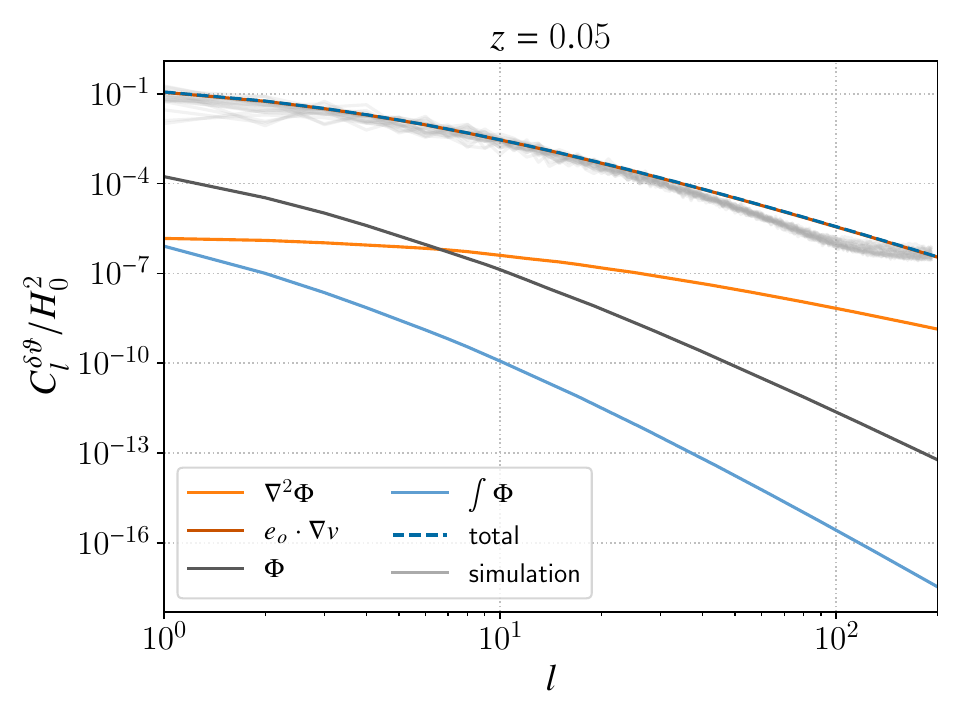}%
    \includegraphics[width=0.4\linewidth]{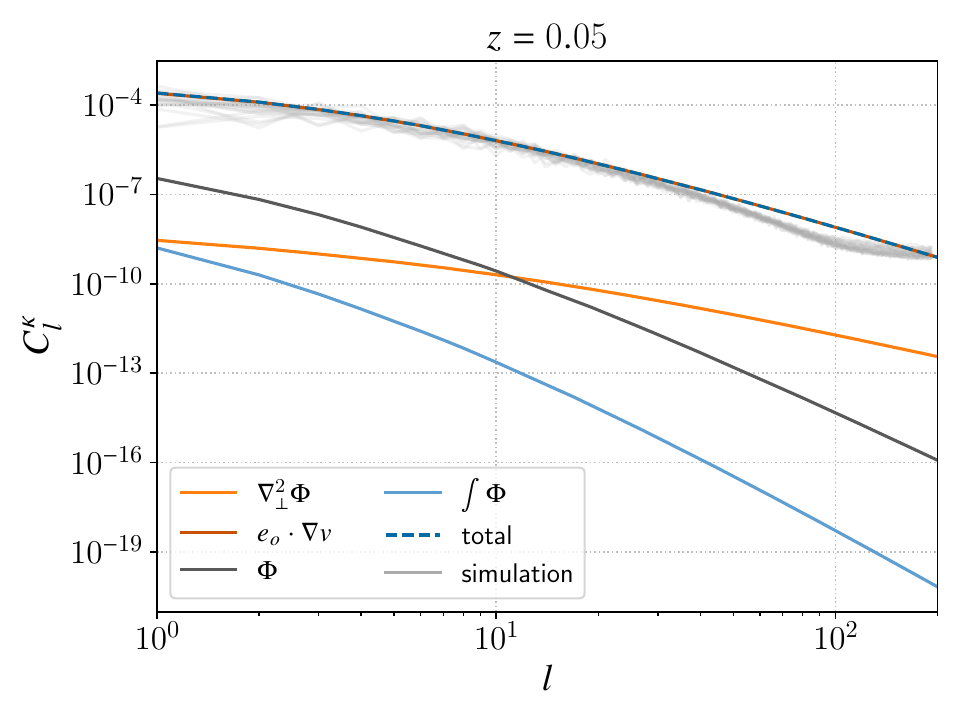}
    \includegraphics[width=0.4\linewidth]{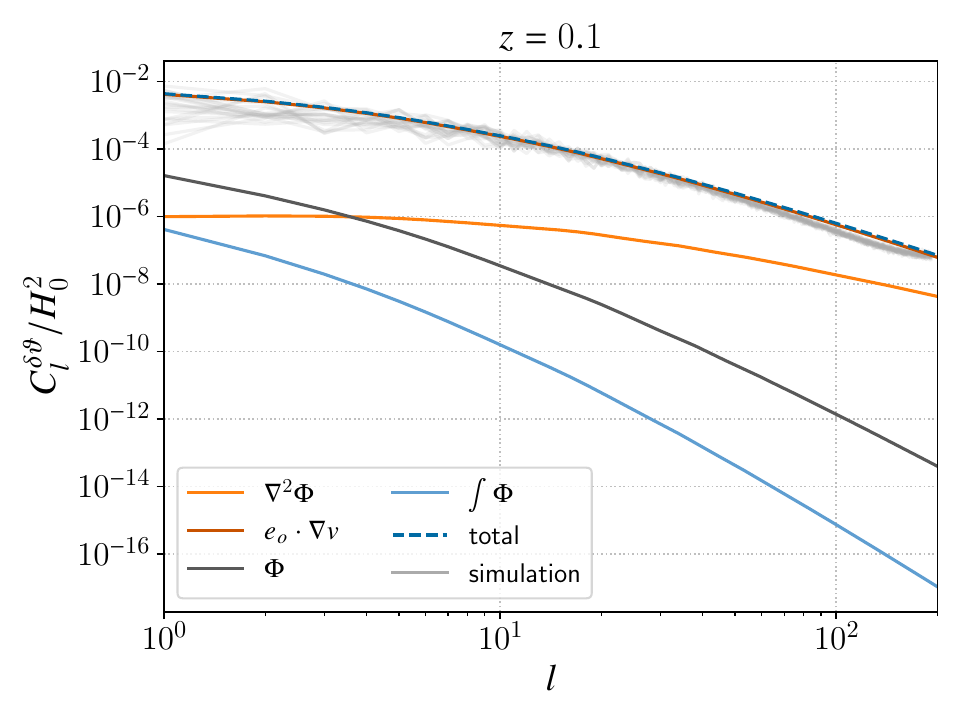}%
    \includegraphics[width=0.4\linewidth]{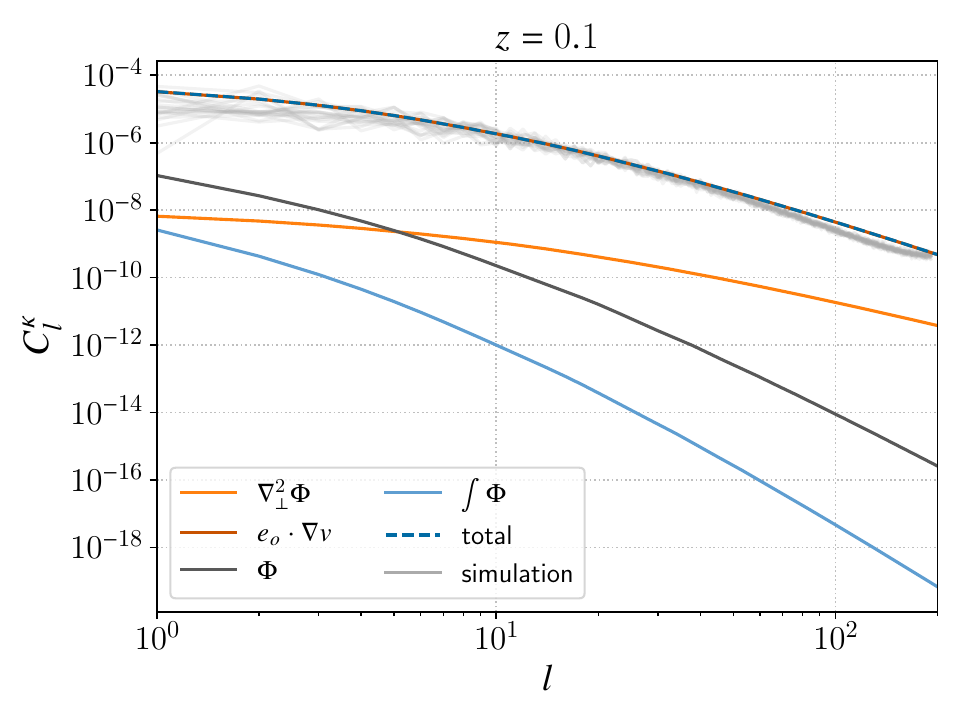}
    \caption{Comparison of the angular power spectra $C_l$ for the fluctuations to the optical expansion scalar $\vartheta$ (left) and the fluctuations to the weak lensing shear $\kappa$ (right) between predictions from linear perturbation theory and results from an N-body simulation. Theoretical spectra are with a Dirac window function. We show the individual contributions to the power spectra as well as the total spectrum.}
    \label{fig:nonlinear}
\end{figure*}

In this subsection, we compare our analytical, linear results with those obtained from direct ray tracing through an N-body simulation. The simulation is run with \texttt{gevolution}\footnote{\url{https://gevolution-code.net}} \cite{Adamek2016_Nat} with initial conditions set at $z = 49$. The simulation contains $2048^3$ particles and has a co-moving box size of $1024$Mpc/h. The cosmological parameters are as summarized earlier in this section.
\newline\indent
To compute $\vartheta$, we wrote a custom ray tracer that reads hdf5 \texttt{gevolution} snapshots of the density field, velocity field and scalar metric perturbation $\Phi$. The ray tracer uses linear interpolation between and within snapshots and computes derivatives through finite differences using \cite{Amaro2024}. The ray tracer solves the geodesic equations \eqref{eq:null_geodesic} backward in time and simultaneously solves the transport equation \eqref{eq:jacobi_transport} for the Jacobian $\b{\mathcal{D}}$. Once $\b{\mathcal{D}}$ and its derivative are known along the light path, we calculate $\theta$ using \eqref{eq:rel_D_S}
\begin{align}
    \theta = \mathrm{tr}(\b{\mathcal{S}})&=\mathrm{tr}(\b{\mathcal{D}}'\b{\mathcal{D}}^{-1}) \nonumber \\ 
    &=\frac{1}{2}\frac{ \mathcal{D}_{11}'\mathcal{D}_{22} + \mathcal{D}_{22}'\mathcal{D}_{11} - \mathcal{D}_{12}'\mathcal{D}_{21} - \mathcal{D}_{21}'\mathcal{D}_{12} }{\mathcal{D}_{11}\mathcal{D}_{22} - \mathcal{D}_{12}\mathcal{D}_{21}}\;,
\end{align}
where primes denote differentiation along the light path, $\D_\Lambda$. We also compute the power spectra for the weak lensing convergence given by 
\begin{align}
    \kappa = 1-\frac{1}{2}\frac{\mathcal{D}_{11}+\mathcal{D}_{22}}{\bar{D}_A}\;, 
\end{align}
which at lowest order in perturbation theory is exactly equal to the relative fluctuations in the angular diameter distance \eqref{eq:kappa=-deltaDA}.
\\ \\
The ray tracer uses the metric implemented in \texttt{gevolution}, but considering only scalar perturbations and neglecting gravitational slip, i.e. assuming $\Phi=\Psi$. The simulations themselves evolve all of the metric perturbations, but the gravitational slip and vector and tensor perturbations are much smaller than $\Phi$ \cite{Adamek2016_Nat}. Neglecting these sub-dominant effects and only including scalar perturbations, the line element implemented in \texttt{gevolution} and used for the ray tracer reduces to 
\begin{align}
    ds^2 = a^2(-\exp{(2\Phi)}d\eta^2 + \exp{(-2\Phi)}\delta_{ij}dx^i dx^j)\;.
\end{align}
We considered 20 randomly placed present-time observers who each obtain light from the entire sky using a \texttt{HEALPix}\footnote{\url{http://healpix.sourceforge.net}} \cite{HEALPix} grid with $N_\mathrm{side} = 64$ so that the total number of rays per observer is $64\times12\times12$. The resulting power spectra (obtained with \texttt{healpy} \cite{Zonca2019_healpy}) are shown in Fig.~\ref{fig:nonlinear}. In the figures we also show our theoretical results from linear perturbation theory. We now show only the dominant terms according to \eqref{eq:theta_dominant_terms} and \eqref{eq:DA_dominant_terms}. The redshift of the sky maps produced with our ray tracer has accuracy $\sim 10^{-4}$. When comparing the simulation results with our perturbative results we therefore use a Dirac window function. We show results for $z=0.05$ and $0.1$.

We note that with the Dirac window function, the contribution from the peculiar velocity dominates the power spectrum for all $l$. The lensing contribution ($\nabla^2\Phi$) is even subdominant on the largest scales we consider, with $l\sim 100$. This is in clear contrast to the power spectra obtained with a Gaussian window function  (see Fig.\ref{fig:cl_gauss_lowz}).

The spectra obtained from the simulations overall agree well with the linear expectations on intermediate scales, i.e. for $10\lesssim l\lesssim 50$. On larger scales with $l\lesssim 10$, there is a large amount of scatter between the different observers, as is expected due to cosmic variance. We note that more observers seem to measure less power than predicted by the linear theory for these $l$. However, with only 20 observers, it is unclear if this is a genuine effect. On the very smallest scales we show ($l\gtrsim 50$), the non-linear results all lie below the prediction from  theory. There are two possible explanations for this:

As seen from the figures, at these low redshifts, both $\delta\vartheta$ and $\kappa$ almost exclusively probe the velocity field since all other contributions to the power spectra are highly sub-dominant and entirely negligible. The non-linear power spectra of the velocity field can in general be split into a contribution from the divergence and the curl of the velocity field. As non-linear structures form, the divergence power spectrum typically falls below linear predictions and is overtaken by the vorticity power spectrum \cite{Hahn2015, Jelic-Cizmek2018, Lepori_2026}. This happens because angular momentum conservation stops the inflow of matter into over-densities, transforming the inflow into rotation. This is a non-linear effect which is thus not captured by the linear power spectra. Instead, both the linear power spectrum for $\vartheta$ and $\kappa$ contain only the gradient of the velocity field and thereby by definition no rotation. The non-linear spectra could therefore be expected to fall below the linear prediction. This is not necessarily the case though; when observing on the sphere, the gradient and rotation contributions to the velocity field mix, and the divergence power spectrum can in principle rise above linear predictions as found in \cite{Oestreicher_2025_Position_Drift} when considering the position drift. 

It is also possible that the reduced power in the non-linear spectra is simply an artefact from the finite resolution of our simulations.
A dedicated set of simulations and analyses are necessary to conclude which of the above possibilities is/are the correct one(s). As our goal here is merely to verify our analytical results and compare them with simulations, we leave a detailed investigation of the non-linear effects to future studies.

\section{Conclusion and Outlook}
\label{sect:conclusions}
It was recently pointed out in \cite{Koksbang_2026_PRD} that Sachs' optical expansion scalar can be constrained observationally by combining observations of the angular diameter distance $D_A$, and the effective observed expansion rate $\mathfrak{H}$. Motivated by this, we here introduced a new observable, $\vartheta$, defined as Sachs' optical expansion scalar divided by the frequency of the observed light, i.e. $\vartheta\equiv\theta/\omega_o$. We derived the expression for $\vartheta$ in a perturbed, flat FLRW spacetime and obtained the corresponding angular power spectra. This provides the first complete perturbative framework for studying the statistical properties of $\vartheta$.
\newline\indent
We compared the power spectra of $\vartheta$ at different redshifts with those of $\delta D_A/\bar{D}_A$. The comparison shows no striking immediate advantage of studying $\vartheta$ over $\delta D_A$. However, this comparison should not be interpreted as indicating that $\vartheta$ contains no additional information. In particular, as $\vartheta$ is a scaled version of the optical expansion along the observer's past light cone, it provides access to a fundamental geometric property of light propagation. Furthermore, the presented initial investigation does provide hints that $\vartheta$ may be better for certain constraints. For instance, at $z\gtrsim0.5$ and lower $l$, the power spectrum of $\delta\vartheta$ remains sensitive to the peculiar velocity and a contribution from the relativistic potentials. Such terms are less significant for the power spectrum of $\delta D_A$, indicating that $\delta\vartheta$ might allow observational access to them where $\delta D_A$ alone does not. $\delta\vartheta$ overall has a more intricate dependence on the metric fluctuations, in particular when gravitational slip is non-negligible. This might allow probing modifications to general relativity more easily.
\newline\indent
One immediate application we anticipate concerns the low multipoles. We demonstrated that the dipole of $\vartheta$ exhibits a significantly more complex redshift dependence than the dipoles of $D_A$ and $\mathfrak{H}$ individually. Because $\vartheta$ is constructed by measurements of $D_A$ and $\mathfrak{H}$, its dipole and other large-scale anisotropies constitute powerful internal consistency checks between the data sets. In particular, if the observed dipoles of $D_A$ and $\mathfrak{H}$ exhibit the expected redshift dependence while $\vartheta$ does not, this would indicate the presence of unaccounted-for systematics or a breakdown of the assumptions used when reconstructing $\vartheta$. Understanding such consistency relations is particularly important given the ongoing tensions between different dipole measurements \cite{dipole} (but see e.g. also \cite{dipole_fine}).
\newline\indent
More broadly, as the optical expansion scalar is one of the fundamental quantities describing light propagation in the universe, we expect that the principal scientific usefulness of $\vartheta$ is likely to emerge beyond the regime considered in this work, namely by providing probes of beyond-FLRW symmetries and modified gravity. The formalism presented here establishes the basic theoretical foundation required for investigating these possibilities.
\newline\newline
Measurements of the quantities needed to construct $\vartheta$ are rapidly improving. Existing BAO data and supernovae compilations of a few thousand supernovae already permit model-independent reconstructions of $D_A, D_A', D_A'', \mathfrak{H}$, and $\mathfrak{H}'$ \cite{Koksbang_2026_PRD}. Upcoming data from e.g. LSST will increase the number of supernovae available for cosmology by more than a factor of $100$ \cite{LSST_2009_Science_Book} and at the same time, surveys such as Euclid, LSST and DESI are expected to provide increasingly precise measurements of the effective observed expansion rate $\mathfrak{H}$. As observations enter this high-precision regime, it is becoming timely to develop the theoretical framework to combine these measurements and exploit the information encoded in observables such as $\vartheta$.

\acknowledgments
The project was funded by VILLUM FONDEN, grant VIL53032. The authors thank Giulia Muco, Julian Adamek, Ruth Durrer and Chris Clarkson for correspondence on various aspects of the work and Chris Clarkson for valuable feedback on the initial draft of the manuscript. The simulations used in this paper were run using the UCloud interactive HPC system managed by the eScience Centre at the University of Southern Denmark.
\newline\newline
{\bf Author contribution statement}
The majority of the technical work was carried out by AO who performed all the analytical analyses and wrote a dedicated numerical code for computing linear power spectra. SMK conceived the original idea for the project and wrote a dedicated ray tracing code for \texttt{gevolution} output for obtaining the power spectra of the optical expansion rate and convergence. Analyses of results and the writing of the manuscript were joint struggles.

\appendix

\section{Geometric Quantities}
\label{appx:geometry}
In this appendix we briefly recap some well known literature results in cosmological perturbation theory that are required for the main calculations presented in the main text. We will need the Christoffel symbols, Ricci scalar, Ricci lensing scalar, perturbations to the wave vector, observed frequency and redshift for the metric whose line element reads
\begin{equation}
	\label{eq:conf_metric_appA}
	{\d s}^2 =\left[-(1+2\Psi)\d\eta^2+(1-2\Phi)\delta_{ij}\d x^i\d x^j\right]\;.
\end{equation}

\subsection{Christoffel Symbols and Ricci Tensor}
The Christoffel symbols for $g_{\mu\nu}$ are \cite{Baumann_2022_Cosmology}
\begin{align}
	\label{eq:christoffels_pertb_minkowski}
	&\Gamma^0_{00}=\dot\Psi\;, \nonumber\\
	&\Gamma^0_{i0}=\p_i\Psi\;, \nonumber\\
	&\Gamma^0_{ij}=-\dot \Phi \delta_{ij}\;, \nonumber\\
	&\Gamma^i_{00}=\delta^{ij}\p_j\Psi \;, \nonumber\\
	&\Gamma^i_{0j}=-\dot\Phi\delta_j^i\;, \nonumber\\
	&\Gamma^i_{jm}=-\delta^i_m\p_j\Phi-\delta_j^i\p_m\Phi+\delta_{jm}\delta^{in}\p_n\Phi\;.
\end{align}
From these one can calculate the components of the Ricci tensor, which are 
\begin{align}
	\label{eq:ricci_tensor_pertb_minkowski}
	& R_{00}=3\ddot\Phi+\nabla^2\Psi\;, \nonumber \\
	& R_{i0}=2\p_i\dot\Phi\;, \nonumber \\
	& R_{ij}= \delta_{ij}\left[-\ddot\Phi+ \nabla^2\Phi\right]+\p_i\p_j\left(\Phi-\Psi\right)\;.   
\end{align}

\subsection{Ricci Lensing Scalar}
We also need the Ricci Lensing scalar. In our notation, $g_{\mu\nu}$ is the perturbed Minkowski spacetime and as such, all of the corresponding Christoffel symbols and components of the corresponding Ricci tensor vanish at background level in perturbation theory. This can also be seen from \eqref{eq:christoffels_pertb_minkowski}, \eqref{eq:ricci_tensor_pertb_minkowski}. We therefore have 
\begin{equation}
	\bar{\mathcal{R}}=-\frac{1}{2}\bar k^\mu \bar k^\nu \bar R_{\mu\nu} = 0\;.
\end{equation}
For the first order perturbations we have
\begin{align}
	\delta\mathcal{R}=&\,-\frac{1}{2}\bar k^\mu\bar k^\nu \delta R_{\mu\nu}\;.
\end{align}
Introducing $n^i=\bar k^i/\bar k^0$, setting $\bar k^0=1$ and using $n^i n_i=1$ we find
\begin{align}
	\delta\mathcal{R}=&\, -\frac{1}{2}\left[ \delta R_{00}+2 n^i \delta R_{0i}+n^i n^j \delta R_{ij}\right]\;, \nonumber \\
	=&\, -\frac{1}{2}\left[2\ddot\Phi+\nabla^2(\Phi+\Psi)+4 n^i \p_i\dot\Phi+n^i n^j \p_i\p_j\left(\Phi-\Psi\right)\right]\;.
\end{align}
We can combine a number of these terms, using the definition of the directional derivate along the background path
\begin{equation}
	\label{eq:replace_lambda_appendix}
	\dl = \bar k^\mu \p_0 = \p_0 +n^i\p_i
\end{equation}
and using that $\dl^2=\bar k^\mu\bar k^\nu\p_\mu\p_\nu$, because the Christoffel symbols vanish at the background level. A simplified expression can then be written as 
\begin{align}
	\delta\mathcal{R}=&\, -\frac{1}{2}\left[\nabla^2(\Phi+\Psi)-n^i n^j \p_i\p_j\left(\Phi+\Psi\right)+2\dl^2\Phi\right]\;.
\end{align}
It is often convenient to express our results in terms of a derivative perpendicular to the direction in which the wave is travelling. We therefore define
\begin{equation}
	\nabla_\perp \equiv  \nabla -\nabla_\parallel\;, \quad \textrm{where}\quad \nabla_\parallel=\b n (\b n\cdot \nabla)\;.
\end{equation}
The perpendicular Laplacian is \cite{Umeh_2014_DA_second_order_derivation}
\begin{equation}
    \label{eq:nabla_perp}
	\nabla_\perp^2 = \nabla^2-n^i n^j\p_i\p_j+\frac{2}{r}n^i\p_i\;,
\end{equation}
where $r$ is the background co-moving distance $r=\eta_o-\eta$. With this, it is possible to write 
\begin{align}
	\delta\mathcal{R}=&\, -\frac{1}{2}\left[\nabla_\perp^2(\Phi+\Psi)+2\dl^2\Phi-\frac{2}{r}n^i\p_i(\Phi+\Psi)\right]\;.
\end{align}
Replacing the remaining spatial derivatives using \eqref{eq:replace_lambda_appendix}, this can also be written as
\begin{align}
    \label{eq:ricci_lensing}
	\delta\mathcal{R}=&\, -\frac{1}{2}\left[\nabla_\perp^2(\Phi+\Psi)+2\dl^2\Phi-\frac{2}{r}\dl(\Phi+\Psi)+\frac{2}{r}(\dot\Phi+\dot\Psi)\right]\;.
\end{align}

\subsection{Perturbations to the Wave Vector}
The perturbations to the wave vector are found from the null geodesic equation
\begin{equation}
    \label{eq:null_geodesic}
    \d_\Lambda k^\mu = -\Gamma^\mu_{\alpha\beta}k^\alpha k^\beta\;.
\end{equation}
At the background level all the Christoffel symbols vanish \eqref{eq:christoffels_pertb_minkowski} and we just have (using \eqref{eq:deriv_pertb})
\begin{equation}
    \label{eq:null_geodesic_eq_bar_k}
	\dl\bar k^\mu=0\;,
\end{equation}
telling us that $\bar k^\mu$ is constant along $\lambda$. \eqref{eq:deriv_pertb}. For the first order perturbations we have 
\begin{equation}
	\dl\delta k^\mu+\delta k^\nu\p_\nu\bar k^\mu=-\delta\Gamma^\mu_{\alpha\beta}\bar k^\alpha \bar k^\beta\;, 
\end{equation}
where we again used \eqref{eq:deriv_pertb} to replace $\dL$. Because we are solving the equation along the background light ray, on which $\bar k^\mu$ is constant, the partial derivatives $\p_\nu\bar k^\mu$ have to vanish and we have 
\begin{equation}
	\dl\delta k^\mu=-\delta\Gamma^\mu_{\alpha\beta}\bar k^\alpha \bar k^\beta\;.
\end{equation}
We will only need the time component in the main text. Inserting the Christoffel symbols \eqref{eq:christoffels_pertb_minkowski} and setting $\bar k^0=1$ we find that the time component can be written as
\begin{align}
	\dl\delta k^0&=-\dot\Psi-2 n^i\p_i\Psi+\delta_{ij}n^i n^j \dot\Phi\;.
\end{align}
We can replace the spatial gradients $n^i\p_i$ with time and directional derivatives using \eqref{eq:replace_lambda_appendix}. Doing this and further more noting that $n^i n_i=1$, leaves us with 
\begin{align}
    \label{eq:bar_d_lambda_deltak0}
	\dl\delta k^0&=-2\dl\Psi+(\dot\Phi+\dot\Psi)\;, 
\end{align}
which after integration becomes 
\begin{equation}
    \label{eq:pertb_k0_lambda}
	\delta k^0(\lambda)=\delta k^0(\lambda_o)-2[\Psi]_\lo^\l+\int_\l\d\l'\,(\dot\Phi+\dot\Psi)\;.
\end{equation}

\subsection{Perturbations to the Frequency and Redshift}
In order to calculate the perturbations to the frequency we first need the 4-velocity. to first order in perturbation theory, the velocity is 
\begin{equation}
    u^\mu = (1-\Psi, v^i)\;, 
\end{equation}
which one can find simply by using $u_\mu u^\mu=-1$. Here, $v^i=u^i/u^0=\d x^i/\d\eta$ is the coordinate/peculiar velocity, which is a pure perturbation and zero in the background. Using this result together with the definition of the observed frequency \eqref{eq:def_omega}, we see that 
\begin{align}
    \bar\omega=&\bar k^0=1\;, \\
    \label{eq:domega=dk0+psi-nv}
    \delta\omega=&\delta k^0+\Psi-v_in^i\;.
\end{align}
From the definition of the redshift \eqref{eq:def_redshift} and our result for the frequency one sees that the former has perturbations 
\begin{align}
    \label{eq:pertb_z}
    \delta z &= \delta\omega-\delta\omega_o \nonumber \\
    &=-n^i[ v_i]^{\l}_{\lo}-[\Psi]_{\lo}^{\l}+\int_\l\,(\dot\Phi+\dot\Psi)\;.
\end{align}

\section{Alternative Expression for $\delta\vartheta$}
In this appendix we rewrite the $\nabla_\perp^2$ term in the expression \eqref{eq:theta_final_result} in terms of $\nabla^2$. From the definition \eqref{eq:nabla_perp} we have 
\begin{align}
    \frac{1}{2r^2}\int_r\,&r'^2\nabla_\perp^2(\Phi+\Psi) \nonumber \\
    &=\frac{1}{2r^2}\int_r\,r'^2\left(\nabla^2-n^i n^j\p_i\p_j+\frac{2}{r}n^i\p_i\right)(\Phi+\Psi)\;.
\end{align}
Using that $\d_\lambda=\bar k^\mu\p_\mu$ and $\dl^2=\bar k^\mu \bar k^\nu \p_\mu \p_\nu$ this may be rewritten as 
\begin{align}
    \frac{1}{2r^2}\int_r\,r'^2\left(\nabla^2-\p_0^2-\d_\lambda^2+2\d_\lambda\p_0+\frac{1}{r'}(\d_\lambda-\p_0)\right)(\Phi+\Psi)\;.
\end{align}
With $r=\l-\lo$ and $\d r=-\d\lambda$ this can be partially integrated leading to
\begin{align}
    &\frac{1}{2r^2}\int_r\,r'^2\left(\nabla^2-\p_0^2\right)(\Phi+\Psi)+\frac{1}{r^2}\int_r\,r'(\dot\Phi+\dot\Psi) \nonumber \\
    &+\frac{1}{2}\dl(\Phi+\Psi)-(\dot\Phi+\Psi)\;.
\end{align}
Again using $\dl=\bar k^\mu\p_\mu$ and replacing $n^i=-e^i_o$ according to \eqref{eq:ei=-ni} this becomes 
\begin{align}
    &\frac{1}{2r^2}\int_r\,r'^2\left(\nabla^2-\p_0^2\right)(\Phi+\Psi)+\frac{1}{r^2}\int_r\,r'(\dot\Phi+\dot\Psi) \nonumber \\
    &-\frac{1}{2}(\dot\Phi+\Psi)-\frac{1}{2}e^i_o\p_i(\Phi+\Psi)\;,
\end{align}
which can now be used to replace the $\nabla^2_\perp$ term in \eqref{eq:theta_final_result} leading to the result \eqref{eq:theta_final_v2}.

\label{appx:rewriting_theta}

\section{Angular Power Spectra}
\subsection{Derivation}
\label{appx:powerspectra}
In this appendix, We derive the angular power spectra for any observable $O(\b e_o, z)$ that can be written as 
\begin{equation}
    O(\b e_o, z) =\sum_i F_i(\b e_o, z)\;, \;\; \textrm{where} \;\; F_i \in \left\{f\;,\b e_o\cdot\nabla f\;, \nabla^2_\Omega f \right\}\;.
\end{equation} 
We here distinguished between terms with different angular dependency which will lead to different types of contributions to the angular power spectrum. The function $f$ will usually be one of the perturbation variables $f_i\in\{\Phi, \Psi, v^i\}$, but for now we apply no restrictions to it.
\newline\indent
Since $O(\b e_o, z)$ is a function on the sphere, we can expand it into spherical harmonics 
\begin{equation}
	O(\b e_o,z) = \sum_{l,m}^\infty a_{lm}^O(z)Y_{lm}(\b e_o)\;,
\end{equation} 
where
\begin{equation}
	a_{lm}^O(z) = \int \d\Omega_{\b e_o} O(\b e_o, z)Y_{lm}^*(\b e_o)\;.
\end{equation}
Since this expression is linear in $O$, we can now study all the contributions to the sum separately and put everything back together in the end. We start with the simplest  contribution
\begin{equation}
	a_{lm}^f(z) = \int \d\Omega_{\b e_o} f(\b e_o, z)Y_{lm}^*(\b e_o)\;.
\end{equation}
Making use of the 3D Fourier Transform, we make the decomposition 
\begin{align}
	f(\b x,z) &= \frac{1}{(2\pi)^3}\int \d^3 k\; f (\b k, z) \e^{-i \b k\b x}\nonumber \\ 
    &= \frac{1}{(2\pi)^3}\int \d^3 k\; f (\b k, z) \e^{-i r \b k\b e_o}\;, 
\end{align}
where we in the second line introduced the co-moving distance $r$ to our source. With this we can now write
\begin{equation}
	a_{lm}^f(z) = \frac{1}{(2\pi)^3}\int \d\Omega_{\b e_o}Y_{lm}^*(\b e_o)\int \d^3 k\; f (\b k, z) \e^{-i r \b k\b e_o}\;.
\end{equation}
Next, we expand the exponential into plane waves
\begin{equation}
    \label{eq:e_in_Ylm}
	\e^{-i r \b k\b e_o} = 4\pi \sum_{l=0}^\infty\sum_{m=-l}^l (-i)^l j_l(kr) Y_{lm}(\b e_o)Y_{lm}^*(\hat{\b k})\;,
\end{equation}
 where $\hat {\b k}$ is the unit vector in the direction $\b k$. The scalar $k$ is the absolute value of $\b k$ and $j_l$ are the spherical Bessel functions of the first kind. With this we have 
\begin{align}
	a_{lm}^f(z) =&\, \frac{1}{2(\pi)^2}\int \d\Omega_{\b e_o}Y_{lm}^*(\b e_o)\int \d^3 k\; f (\b k, z) \nonumber \\
    &\,\sum_{l'=0}^\infty\sum_{m'=-l'}^{l'} (-i)^{l'} j_{l'}(kr) Y_{l'm'}(\b e_o)Y_{l'm'}^*(\hat{\b k})\;.
\end{align}
Using the orthonormality of the spherical Bessel functions
\begin{equation}
	\int \d\Omega_{\b e_o}Y_{lm}^*(\b e_o)Y_{l'm'}(\b e_o) = \delta_{mm'}\delta_{ll'}
\end{equation}
and summing over the two delta functions leaves us with 
\begin{equation}
	a_{lm}^f(z) = \frac{(-i)^l}{2(\pi)^2}\int \d^3 k\; f (\b k, z) j_{l}(kr)Y_{lm}^*(\hat{\b k})\;.
\end{equation}
Let us now assume that we can in some way relate our function to an initial random variable that we take to be the Bardeen potential $\Phi$, i.e.
\begin{equation}
    f(\b k, z)=T_f(k,z)\Phi_\textrm{in}(\b k)\;,
\end{equation}
where $T_f(k,z)$ is a function that depends only on the absolute value of the wave vector $k$ and no longer on direction. For convenience, we define 
\begin{equation}
    \mathcal{T}_f(k,z,l)= T_f(k,z)j_l(k,r(z)).
\end{equation}
With this definition in place, we find 
\begin{align}
	\langle a_{lm}^f a_{l'm'}^{*f}\rangle =&\, \frac{(-i)^l(i)^{l'}}{4\pi^4}\int \d^3k\;\int \d^3 k'\; \mathcal{T}_f(k,z,l) \nonumber \\ 
    &\,\mathcal{T}_f(k',z,l')Y_{lm}^*(\hat{\b k})Y_{l'm'}(\hat{\b k}')\langle \Phi_\textrm{in}(\b k)\Phi_\textrm{in}^*(\b k')\rangle\;.
\end{align}
Using the definition of the power spectrum 
\begin{equation}
	\langle \Phi_\textrm{in}(\b k)\Phi_\textrm{in}^*(\b k')\rangle = (2\pi)^3 \delta (\b k -\b k')\mathcal{P}(k) 
\end{equation}
and integrating over the appearing delta function yields 
\begin{align}
	\langle a_{lm}^f a_{l'm'}^{*f}\rangle =&\, \frac{2(-i)^l(i)^{l'}}{\pi}\int \d^3k\; \mathcal{T}_f(k,z,l) \nonumber \\
    &\,\mathcal{T}_f(k',z,l')Y_{lm}^*(\hat{\b k})Y_{l'm'}(\hat{\b k})\mathcal{P}(k)\;.
\end{align}
Using the orthonormality of the spherical harmonics to carry out the angular integral leaves us with 
\begin{align}
	\langle a_{lm}^f a_{l'm'}^{*\delta}\rangle =&\, \delta_{ll'}\delta_{mm'} \frac{2(-i)^l(i)^{l'}}{\pi}\int \d k k^2\; \mathcal{T}_f(k,z,l)\nonumber \\ 
    &\,\mathcal{T}_f(k',z,l')\mathcal{P}(k)\;.
\end{align}
Comparing with the definition of the angular power spectrum 
\begin{equation}
	\delta_{ll'}\delta_{mm'}C_l = \langle a_{lm} a_{l'm'}^*\rangle\;, 
\end{equation}
we see that 
\begin{equation}
	C_l^f(z) = \frac{2}{\pi}\int \d k k^2\; \mathcal{P}(k)\mathcal{T}_f^2(k,z,l)\;.
\end{equation}
In a real survey we do not observe at one specific redshift but rather in a redshift-bin around a mean redshift. We can account for this finite width of the redshift bin by adding a window function to the definition of $\mathcal{T}_f$ according to
\begin{equation}
	\mathcal{T}^W_f(k,z,l) = \int_0^\infty \d z'\; W(z,z') T_f(k,z',j)\;.
\end{equation}
This expression is valid for ordinary auto-correlations. If we are instead interested in cross correlations between two different functions at two different redshifts, $f(z_1)$ and $g(z_2)$, the $C_l$ become
\begin{equation}
	C_l^{fg}(z_1,z_2) = \frac{2}{\pi}\int \d k k^2\; \mathcal{P}(k)\mathcal{T}^W_f(k,z_1,l)\mathcal{T}^W_g(k,z_2,l)\;.
\end{equation}
We now move on to the next type of term, namely the term $\b e_o\cdot \nabla f$ which in Fourier space becomes $-i\b e_o\cdot\b kf$. In this case, we find
\begin{align}
	a_{lm}^{\b e_o\cdot\nabla f}(z) =&\, \frac{1}{(2\pi)^3}\int \d\Omega_{\b e_o}Y_{lm}^*(\b e_o) \nonumber \\ 
    &\,\int \d^3 k\; (-i\b e_o\cdot\b k)f (\b k, z) \e^{-i r \b k\b e_o} \\
    =&\, \frac{1}{(2\pi)^3}\int \d\Omega_{\b e_o}Y_{lm}^*(\b e_o)\nonumber \\ 
    &\,\int \d^3 k\; k f (\b k, z) \p_{rk}\,\e^{-i r \b k\b e_o}\;.
\end{align}
Expanding the exponential into spherical harmonics (using \eqref{eq:e_in_Ylm}) the derivative can be evaluated. Proceeding with the same steps as for the previous type of term, we find that the derivation remains almost exactly the same except that the function $\mathcal{T}_f$ now is
\begin{equation}
	\mathcal{T}_{\b e_o\cdot\nabla f}(k,z,l) = k T_f(k,z) j'_l(k,r(z))\;.
\end{equation}
Here, $j'_l$ denotes a derivative of the Bessel function with respect to its argument.
\\ \\
Lastly, we need terms that contain an angular derivative $\nabla^2_\Omega f$. Following the same steps as before we have the $a_{lm}$ given by
\begin{align}
	a_{lm}^{\nabla^2_\Omega f}(z) &= \frac{1}{(2\pi)^3}\int \d\Omega_{\b e_o}Y_{lm}^*(\b e_o)\nabla^2_\Omega\int \d^3 k\;f (\b k, z) \e^{-i r \b k\b e_o}\;.
\end{align}
Once again expanding the exponential into spherical harmonics using \eqref{eq:e_in_Ylm} we can use the identity $\nabla^2_\Omega Y_{lm}(\b e_o)=-l(l+1)Y_{lm}$ and then proceed as earlier to find that the only thing that has changed compared to the earlier derivations is the function $\mathcal{T}_f$. We therefore find
\begin{equation}
	\mathcal{T}_{\nabla^2_\Omega f}(k,z,l) = -l(l+1)T_f(k,z) j_l(k,r(z))\;.
\end{equation}
With this, we have solved all the cases needed to obtain the power spectrum for the expansion scalar. What remains is to determine the function $T_f$ for each of the terms in $O$. Since only the function $\mathcal{T}_f$ changes for the different $a_{lm}$ and $O$ is just a sum of all the different terms, it is straightforward to see that
\begin{equation}
    C_l^{O_1 O_2}(z_1, z_2)= \frac{2}{\pi}\int \d k k^2\; \mathcal{P}(k)\mathcal{T}^W_{O_1}(k,z_1,l)\mathcal{T}^W_{O_2}(k,z_2,l)\;,
\end{equation}
where
\begin{equation}
    \mathcal{T}^W_O=\int\d z'\,W(z,z')\mathcal{T}_O(k,z',l)\;, 
\end{equation}
and
\begin{equation}
    \mathcal{T}_O(k,z,l)=\sum_i\mathcal{T}_{F_i}(k,z,l)\;.
\end{equation}
The different $\mathcal{T}_{F_i}$ are 
\begin{align}
    \mathcal{T}_f(k,z,l) &= T_f(k,z)j_l(k,r(z))\, \\
    \mathcal{T}_{\b e_o\cdot\nabla f}(k,z,l) &= kT_f(k,z) j'_l(k,r(z))\, \\ 
    \mathcal{T}_{\nabla^2_\Omega f}(k,z,l) &= -l(l+1)T_f(k,z) j_l(k,r(z))\;.
\end{align}
We have allowed for two different observables $O_1, O_2$ at two different redshifts, in order to allow for the calculation of cross-spectra as well as auto-spectra.

\subsection{Numerical Evaluation}
\label{appx:power_spectra_numerics}
The integrals appearing in the angular power spectra are highly oscillatory. In order to solve them efficiently we realize that all the transfer functions we consider are related to $T_\Phi$ such that we can separate them as 
\begin{equation}
	\mathcal{T}_{f_i}(k,l,z)=k^n\, T^l_{f_i}(l)\, T^z_{f_i}(z)\,j^d_l(kr)\, T_\Phi^0(k)\;,
\end{equation}
where $n$ is a real number giving the power of $k$ prefactors, $T^l$ depends only on $l$, $T^z$ only on $z$, $d$ indicates the order of the derivative of the Bessel function and we defined $T_\Phi^0(k)\equiv T_\Phi(k, z=0)$. Adding a window function we have
\begin{align}
	\mathcal{T}^W_{f_i}(k,z,l)=&\,k^n\, T^l_{f_i}(l)\, T_\Phi^0(k)\nonumber \\ 
	&\,\int_0^\infty\d r\, W(r(z),r)\,T^z_{f_i}(r)\,j^d_l(kr)\;.
\end{align}
It can happen that we have terms that involve further integrals along the light path
\begin{align}
	\mathcal{T}^W_{\int f_i}(k,z,l) =&\, k^n\, T^l_{f_i}(l)\, T_\Phi^0(k)\int_0^\infty\d r\, W(r(z),r)\nonumber \\ 
	&\, \int_0^{r}\d r'\,T^z_{f_i}(r,r')\,j^d_l(kr')\;, \\
   \mathcal{T}^W_{\iint f_i}(k,z,l) =&\, k^n\, T^l_{f_i}(l)\, T_\Phi^0(k)\int_0^\infty\d r\, W(r(z),r)\nonumber \\
	&\,\int_0^{r}\d r'\,\int_0^{r'}\d r''\,T^z_{f_i}(r,r',r'')\,j^d_l(kr'')\;.
\end{align}
For terms of this type it is now a good idea to invert the order of integration, such that the difficult oscillating integral over the spherical Bessel functions is done last, i.e.
\begin{align}
	\mathcal{T}^W_{\int f_i}(k,z,l) =&\, k^n\, T^l_{f_i}(l)\, T_\Phi^0(k) \int_0^\infty\d r'\,j^d_l(kr') \nonumber \\ 
	&\, \int_{r'}^\infty\d r\,W(r(z),r)T^z_{f_i}(r,r')\;, \\
   \mathcal{T}^W_{\iint f_i}(k,z,l) =&\, k^n\, T^l_{f_i}(l)\, T_\Phi^0(k) \int_0^\infty\d r''\,j^d_l(kr'') \nonumber \\ 
	&\,\int_{r''}^\infty\d r'\,\int_{r'}^\infty\d r\,W(r(z),r)T^z_{f_i}(r,r',r'')\;.
\end{align}
In this form, the inner integrals are easily solved using a standard integrator and the integral over the Bessel functions can be solved using the library \texttt{FFTlog-and-beyond} \cite{Fang_2020_FFTlog} specifically designed to handle this type of integral. With the $\mathcal{T}_{f_i}^W$ obtained, the $C_l$ follow after another integration over $k$. For this final $k$-integral and the two inner distance integrals, we use Simpson's rule.

\bibliography{main}

\end{document}